# Dissipative analysis of linear coupled differential-difference systems with distributed delays

Qian Feng[a], Sing Kiong Nguang[a], Alexandre Seuret[b]

[a] *Department of Electrical and Computer Engineering, The University of Auckland, Auckland 1010, New Zealand*
[b] *LAAS - CNRS, Université de Toulouse, CNRS, UPS, 7 avenue du Colonel Roche, 31077 Toulouse, France*

**Abstract**

In this paper, we present a new method for the dissipativity and stability analysis of a linear coupled differential-difference system (CDDS) with general distributed delays at both state and output. More precisely, the distributed delay terms under consideration can contain any $\mathbb{L}^2$ functions which are approximated via a class of elementary functions which includes the option of Legendre polynomials. By using this broader class of functions compared to the existing Legendre polynomials approximation approach, one can construct a Liapunov-Krasovskii functional which is parameterized by non-polynomial functions . Furthermore, a novel generalized integral inequality is also proposed to incorporate approximation error in our stability (dissipativity) conditions. Based on the proposed approximation scenario with the proposed integral inequality, sufficient conditions determining the dissipativity and stability of a CDDS are derived in terms of linear matrix inequalities. In addition, several hierarchies in terms of the feasibility of the proposed conditions are derived under certain constraints. Finally, several numerical examples are presented in this paper to show the effectiveness of our proposed methodologies.

## 1. Introduction

Coupled differential-functional equations (CDFEs), which are mathematically related to time-delay systems Briat (2014), can characterize a broad class of models concerning delay or propagation effects Rsvan (2006). CDESs are able to model systems such as standard or neutral time-delay systems or certain singular delay systems Gu & Niculescu (2006). For more information on the topic of CDFEs, see Gu & Liu (2009); Karafyllis *et al.* (2009) and the references therein.

Over the past decades, a series of significant results on the stability of CDFEs Pepe (2005); Pepe *et al.* (2008) has been proposed based on the approach of constructing Liapunov-Krasovskii functionals. In particular, the idea of complete Krasovskii functional of linear time-delay systems Briat (2014) has been extended in to formulate a complete functional for a linear coupled differential-difference system (CDDS)[1] Gu & Liu (2009), which may be constructed numerically Li (2012) via semidefinite programming. To the best of our knowledge, however, no results have been proposed in the reviewed publications on linear CDDSs with non-trivial (non-constant) distributed delays. Generally speaking, analyzing distributed delays may require much more efforts due to the complexities induced by different types of distributed delay kernels. For the latest existing time domain based results in connection with distributed delays, see Münz *et al.* (2009); Fridman & Tsodik (2009); Goebel *et al.* (2011); Kharitonov (2012); Seuret *et al.* (2015); Feng & Nguang (2016b).

*Email addresses:* qfen204@aucklanduni.ac.nz (Qian Feng), sk.nguang@auckland.ac.nz (Sing Kiong Nguang), aseuret@laas.fr (Alexandre Seuret)

[1] A CDDS can be considered as a special case of the systems characterized by CDFEs

In Seuret *et al.* (2015), an approximation scheme is proposed to deal with $\mathbb{L}^2$ continuous distributed delay terms based on the application of Legendre polynomials. Although only the situation of having one or two distributed delay kernels are considered in Seuret *et al.* (2015), the stability conditions derived in Seuret *et al.* (2015) are highly competent and exhibit a pattern of hierarchical feasibility enhancement with respect to the degree of the approximating Legendre polynomials. In this paper, we propose a new approach generalizing the results in Seuret *et al.* (2015). Unlike the approximation scheme in Seuret *et al.* (2015) where approximation is solely attained by the application of Legendre orthogonal polynomials, our proposed approximation solution is based on a class of elementary functions (this including the case of Legendre polynomials or trigonometric functions). The proposed methodology provides a unified solution which can handle the situations that multiple distributed matrix kernels are approximated individually over two different integration intervals with general matrix structures. Furthermore, unified measures concerning approximation errors are formulated via a matrix framework and these measures are included by our proposed stability and dissipativity condition.

In this paper, we propose solutions for the dissipativity and stability analysis of a linear CDDS with distributed delays at both the states and output equation. Specifically, the distributed delay kernels considered can be any $\mathbb{L}^2$ function and the kernel functions are approximated by a class of elementary functions. Many existing models with delays, such as the ones in Münz *et al.* (2009); Gu & Liu (2009); Li (2012); Seuret *et al.* (2015); Feng & Nguang (2016b) are the special cases of the considered system model in this paper. Meanwhile, analysis of the behavior of the approximation errors is presented by using matrix representations which generalize the existing results in Seuret *et al.* (2015). Furthermore, a quadratic supply function is also considered for the dissipative analysis. To incorporate the approximation errors into the optimization constraints for dissipativity and stability analysis, a general integral inequality is derived which introduces error related terms into its lower bound. By constructing a Krasovskii functional with the assistance of this inequality, sufficient conditions which ensure dissipativity and asymptotic (exponential) stability can be derived in terms of linear matrix inequalities. The proposed conditions are further proved to have a hierarchical feasibility enlargement if only orthogonal functions are chosen to approximate the distributed delay kernels, which can be considered as a generalization of the result in Seuret & Gouaisbaut (2015). Finally, several numerical examples are given to demonstrate the effectiveness and capacity of the proposed methodologies.

## Notation

Throughout this paper: we use $\|\mathbf{x}\|_q = \left(\sum_{i=1}^{n} |x_i|^q\right)^{\frac{1}{q}}$ and $\|f(\cdot)\|_p = \left(\int_{\mathcal{X}} |f(\tau)|^p \mathsf{d}\tau\right)^{\frac{1}{p}}$ and $\|\boldsymbol{f}(\cdot)\|_p = \left(\int_{\mathcal{X}} \|\boldsymbol{f}(\tau)\|_2^p \mathsf{d}\tau\right)^{\frac{1}{p}}$ to denote the norms associated with $\mathbb{R}^n$ and the Lebesgue functions space $\mathbb{L}^p(\mathcal{X} \,⨾\, \mathbb{R})$ and $\mathbb{L}^p(\mathcal{X} \,⨾\, \mathbb{R}^n)$, where $\mathcal{X} \subseteq \mathbb{R}$, respectively. In addition, $\widehat{\mathbb{L}}^p(\mathcal{X} \,⨾\, \mathbb{R}^n)$ contains locally integrable Lebesgue measurable functions with reference to $\mathbb{L}^p(\mathcal{X} \,⨾\, \mathbb{R}^n)$. $\mathbf{Sy}(X) := X + X^\top$ is the sum of a matrix with its transpose. The standard gamma function is denoted by $\gamma(\cdot)$. A column vector containing a sequence of objects is defined as $\mathbf{Col}_{i=1}^{n} x_i := \left[\mathbf{Row}_{i=1}^{n} x_i^\top\right]^\top = \left[x_1^\top \cdots x_i^\top \cdots x_n^\top\right]^\top$. In addition, we define $\mathbf{Col}_{i=1}^{n} = []$ when $n < 1$, where $[]$ is an empty matrix with an appropriate column dimension based on specific contexts. $*$ is applied to denote $[*]YX = X^\top YX$ or $X^\top Y[*] = X^\top YX$. $\mathsf{O}_{n\times n}$ denotes a $n \times n$ zero matrix with the abbreviation form $\mathsf{O}_n$, whereas $\mathbf{0}_n$ denotes a $n \times 1$ column vector. Furthermore, we use the notations $x \vee y = \max(x, y)$ and $x \wedge y = \min(x, y)$. The diagonal sum of two matrices and $n$ matrices are defined as $X \oplus Y = \mathsf{Diag}(X, Y) = \left[\begin{smallmatrix} X & \mathsf{O} \\ \mathsf{O} & Y \end{smallmatrix}\right]$, $\bigoplus_{i=1}^{n} X_i = \mathsf{Diag}_{i=1}^{n}(X_i)$, respectively. $\otimes$ denotes the Kronecker product. Furthermore, we assume the order of operations concerning matrices to be *matrix (scalars) multiplications* $> \otimes > \oplus > +$. Finally, the notion of empty matrices is applied in this article to facilitate our derivation, whose rules of operations are in line with the definition in Matlab environment.

## 2. Preliminaries and Problem Formulations

The following linear CDDS

$$
\begin{aligned}
\dot{\boldsymbol{x}}(t) &= A_1\boldsymbol{x}(t) + A_2\boldsymbol{y}(t-r_1) + A_3\boldsymbol{y}(t-r_2) + \int_{-r_1}^{0}\widetilde{A}_4(\tau)\boldsymbol{y}(t+\tau)\mathsf{d}\tau + \int_{-r_2}^{-r_1}\widetilde{A}_5(\tau)\boldsymbol{y}(t+\tau)\mathsf{d}\tau \\
&\qquad + D_1\boldsymbol{w}(t) \\
\boldsymbol{y}(t) &= A_6\boldsymbol{x}(t) + A_7\boldsymbol{y}(t-r_1) + A_8\boldsymbol{y}(t-r_2), \qquad t \geq t_0 \\
\boldsymbol{z}(t) &= C_1\boldsymbol{x}(t) + C_2\boldsymbol{y}(t-r_1) + C_3\boldsymbol{y}(t-r_2) + \int_{-r_1}^{0}\widetilde{C}_4(\tau)\boldsymbol{y}(t+\tau)\mathsf{d}\tau + \int_{-r_2}^{-r_1}\widetilde{C}_5(\tau)\boldsymbol{y}(t+\tau)\mathsf{d}\tau \\
&\qquad + C_6\dot{\boldsymbol{y}}(t-r_1) + C_7\dot{\boldsymbol{y}}(t-r_2) + D_2\boldsymbol{w}(t) \\
\boldsymbol{x}(t_0) &= \boldsymbol{\xi} \in \mathbb{R}^n, \qquad \forall\theta \in [-r_2, 0),\ \boldsymbol{y}(t_0+\theta) = \boldsymbol{\psi}(\theta), \quad \boldsymbol{\psi}(\cdot) \in \mathcal{A}\left([-r_2,0) \,;\, \mathbb{R}^\nu\right)
\end{aligned} \tag{1}
$$

with distributed delays is considered in this paper, where $r_2 > r_1 > 0$ and $t_0 \in \mathbb{R}$. The notation $\mathcal{A}\left([-r_2,0) \,;\, \mathbb{R}^\nu\right)$ in (1) stands for

$$
\mathcal{A}\left([-r_2,0) \,;\, \mathbb{R}^\nu\right) := \left\{\boldsymbol{\psi}(\cdot) \in \mathbb{C}\left([-r_2,0) \,;\, \mathbb{R}^\nu\right) :\ \dot{\boldsymbol{\psi}}(\cdot) \in \mathbb{L}^2\left([-r_2,0) \,;\, \mathbb{R}^\nu\right)\ \&\ \|\boldsymbol{\psi}(\cdot)\|_\infty + \|\dot{\boldsymbol{\psi}}(\cdot)\|_2 < +\infty\right\}
$$

where $\|\boldsymbol{\psi}(\cdot)\|_\infty := \sup_{\tau\in\mathcal{X}} \|\boldsymbol{\psi}(\tau)\|_2$ and $\dot{\boldsymbol{\psi}}(\cdot)$ stands for the weak derivatives of $\boldsymbol{\psi}(\cdot)$. Furthermore, $\boldsymbol{x}(t) \in \mathbb{R}^n, \boldsymbol{y}(t) \in \mathbb{R}^\nu$ satisfy (1), and $\boldsymbol{w}(\cdot) \in \widehat{\mathbb{L}}^2([t_0,\infty) \,;\, \mathbb{R}^q)$, $\boldsymbol{z}(t) \in \mathbb{R}^m$ are the disturbance and output of (1), respectively. The size of the state space matrices in (1) are determined by the given dimensions $n; \nu \in \mathbb{N}$ and $m; q \in \mathbb{N}_0$. All the functions in the entries of the matrix valued distributed delay terms $\widetilde{A}_4(\cdot), \widetilde{C}_4(\cdot)$ and $\widetilde{A}_5(\cdot), \widetilde{C}_5(\cdot)$ are the elements of $\mathbb{L}^2\left([-r_1,0] \,;\, \mathbb{R}\right)$ and $\mathbb{L}^2\left([-r_2,-r_1] \,;\, \mathbb{R}\right)$, respectively. Finally, $A_7$ and $A_8$ satisfy

$$
\sup\left\{s \in \mathbb{C} : \det\left(I_\nu - A_7\mathsf{e}^{-r_1 s} - A_8\mathsf{e}^{-r_2 s}\right) = 0\right\} < 0, \tag{2}
$$

which ensures input to state stability for the associated difference equation Gu (2010) of (1).

In order to deal with the distributed delay terms in (1), we first define $\acute{\boldsymbol{f}}(\cdot) \in \mathbb{C}^1\left([-r_1,0] \,;\, \mathbb{R}^d\right)$ and $\grave{\boldsymbol{f}}(\cdot) \in \mathbb{C}^1\left([-r_2,-r_1] \,;\, \mathbb{R}^\delta\right)$ which satisfy the conditions:

$$
\exists! M_1 \in \mathbb{R}^{d\times d},\ \exists! M_2 \in \mathbb{R}^{\delta\times\delta} : \frac{\mathsf{d}\acute{\boldsymbol{f}}(\tau)}{\mathsf{d}\tau} = M_1\acute{\boldsymbol{f}}(\tau) \ \text{ and } \ \frac{\mathsf{d}\grave{\boldsymbol{f}}(\tau)}{\mathsf{d}\tau} = M_2\grave{\boldsymbol{f}}(\tau) \tag{3}
$$

$$
\begin{aligned}
&\exists\acute{\boldsymbol{\phi}}(\cdot) \in \mathbb{C}^1([-r_1,0] \,;\, \mathbb{R}^{\kappa_1}),\ \exists\grave{\boldsymbol{\phi}}(\cdot) \in \mathbb{C}^1([-r_2,-r_1] \,;\, \mathbb{R}^{\kappa_2}),\ \exists! M_3 \in \mathbb{R}^{\kappa_1\times d},\ \exists! M_4 \in \mathbb{R}^{\kappa_2\times\delta} : \\
&\qquad\qquad \frac{\mathsf{d}\acute{\boldsymbol{\phi}}(\tau)}{\mathsf{d}\tau} = M_3\acute{\boldsymbol{f}}(\tau) \ \text{ and } \ \frac{\mathsf{d}\grave{\boldsymbol{\phi}}(\tau)}{\mathsf{d}\tau} = M_4\grave{\boldsymbol{f}}(\tau)
\end{aligned} \tag{4}
$$

$$
\mathbb{S}^d \ni \acute{\mathsf{F}}_d^{-1} = \int_{-r_2}^{-r_1}\acute{\boldsymbol{f}}(\tau)\acute{\boldsymbol{f}}^\top(\tau)\mathsf{d}\tau \succ 0, \quad \mathbb{S}^\delta \ni \grave{\mathsf{F}}_\delta^{-1} = \int_{-r_2}^{-r_1}\grave{\boldsymbol{f}}(\tau)\grave{\boldsymbol{f}}^\top(\tau)\mathsf{d}\tau \succ 0 \tag{5}
$$

$$
\mathbb{S}^{\kappa_1} \ni \acute{\Phi}_{\kappa_1}^{-1} = \int_{-r_1}^{0}\acute{\boldsymbol{\phi}}(\tau)\acute{\boldsymbol{\phi}}^\top(\tau)\mathsf{d}\tau \succ 0, \quad \mathbb{S}^{\kappa_2} \ni \grave{\Phi}_{\kappa_2}^{-1} = \int_{-r_2}^{-r_1}\grave{\boldsymbol{\phi}}(\tau)\grave{\boldsymbol{\phi}}^\top(\tau)\mathsf{d}\tau \succ 0 \tag{6}
$$

where $d; \delta \in \mathbb{N}$, and (6) indicates that the functions in $\acute{\boldsymbol{f}}(\cdot)$, $\grave{\boldsymbol{f}}(\cdot)$, $\acute{\boldsymbol{\phi}}(\cdot)$ and $\grave{\boldsymbol{\phi}}(\cdot)$ are linearly independent in a Lebesgue sense, respectively. See Theorem 7.2.10 in Horn & Johnson (2012) for the explanation of the meaning of (6).

**Remark 1.** The constraint in (3) indicates that the functions in $\acute{\boldsymbol{f}}(\cdot)$, $\grave{\boldsymbol{f}}(\cdot)$ are the solutions of homogeneous differential equations with constant coefficients. (polynomials, exponential, trigonometric functions, etc) Note that the conditions in (4) do not put extra constraints on $\acute{\boldsymbol{f}}(\cdot)$, $\grave{\boldsymbol{f}}(\cdot)$. This is because for any given $\acute{\boldsymbol{f}}(\cdot)$, $\grave{\boldsymbol{f}}(\cdot)$ satisfying (3), the one can always to make the choice of $\acute{\boldsymbol{\phi}}(\tau) = \acute{\boldsymbol{f}}(\tau)$ and $\grave{\boldsymbol{\phi}}(\tau) = \grave{\boldsymbol{f}}(\tau)$ with $M_3 = M_1$ and $M_4 = M_2$ which can satisfy (4).

Now given $\acute{\boldsymbol{f}}(\cdot) \in \mathbb{C}^1\big([-r_1, 0] ; \mathbb{R}^d\big)$ and $\grave{\boldsymbol{f}}(\cdot) \in \mathbb{C}^1\big([-r_2, -r_1] ; \mathbb{R}^\delta\big)$ satisfying (3), one can conclude that for any $\widetilde{A}_4(\cdot); \widetilde{A}_5(\cdot)$ and $\widetilde{C}_4(\cdot); \widetilde{C}_5(\cdot)$ in (1), there exist constant matrices $A_4 \in \mathbb{R}^{n\times(d+\mu_1)\nu}$, $A_5 \in \mathbb{R}^{n\times(\delta+\mu_2)\nu}$, $C_4 \in \mathbb{R}^{m\times(d+\mu_1)\nu}$, $C_5 \in \mathbb{R}^{m\times(\delta+\mu_2)\nu}$ and the functions $\boldsymbol{\varphi}_1(\cdot) \in \mathbb{L}^2([-r_1, 0] ; \mathbb{R}^{\mu_1})$, $\boldsymbol{\varphi}_2(\cdot) \in \mathbb{L}^2([-r_2, -r_1] ; \mathbb{R}^{\mu_2})$ such that

$$\begin{aligned}
&\widetilde{A}_4(\tau) = A_4\left(\begin{bmatrix}\boldsymbol{\varphi}_1(\tau)\\ \acute{\boldsymbol{f}}(\tau)\end{bmatrix}\otimes I_\nu\right), \qquad \widetilde{A}_5(\tau) = A_5\left(\begin{bmatrix}\boldsymbol{\varphi}_2(\tau)\\ \grave{\boldsymbol{f}}(\tau)\end{bmatrix}\otimes I_\nu\right)\\
&\widetilde{C}_4(\tau) = C_4\left(\begin{bmatrix}\boldsymbol{\varphi}_1(\tau)\\ \acute{\boldsymbol{f}}(\tau)\end{bmatrix}\otimes I_\nu\right), \qquad \widetilde{C}_5(\tau) = C_5\left(\begin{bmatrix}\boldsymbol{\varphi}_2(\tau)\\ \grave{\boldsymbol{f}}(\tau)\end{bmatrix}\otimes I_\nu\right)
\end{aligned} \tag{7}$$

$$\int_{-r_1}^{0}\begin{bmatrix}\boldsymbol{\varphi}_1(\tau)\\ \acute{\boldsymbol{f}}(\tau)\end{bmatrix}\begin{bmatrix}\boldsymbol{\varphi}_1^\top(\tau) & \acute{\boldsymbol{f}}^\top(\tau)\end{bmatrix}\mathsf{d}\tau \succ 0, \qquad \int_{-r_2}^{-r_1}\begin{bmatrix}\boldsymbol{\varphi}_2(\tau)\\ \grave{\boldsymbol{f}}(\tau)\end{bmatrix}\begin{bmatrix}\boldsymbol{\varphi}_2^\top(\tau) & \grave{\boldsymbol{f}}^\top(\tau)\end{bmatrix}\mathsf{d}\tau \succ 0 \tag{8}$$

where $\mu_1, \mu_2 \in \mathbb{N}_0$ and (8) indicates that the functions in $\mathbf{Col}\left[\boldsymbol{\varphi}_1(\tau), \acute{\boldsymbol{f}}(\tau)\right]$ and $\mathbf{Col}\left[\boldsymbol{\varphi}_2(\tau), \grave{\boldsymbol{f}}(\tau)\right]$ are linearly independent in a Lebesgue sense, respectively. Thus (7) can be applied to equivalently describe the distributed delay terms in (1). Finally, note that (5) is satisfied if (8) holds.

**Remark 2.** The elements in $\acute{\boldsymbol{f}}(\cdot)$ and $\grave{\boldsymbol{f}}(\cdot)$ in (7) are chosen in view of the functions in $\widetilde{A}_4(\cdot)$, $\widetilde{A}_5(\cdot)$, $\widetilde{C}_4(\cdot)$ and $\widetilde{C}_5(\cdot)$. Note that one can always let $\acute{\boldsymbol{f}}(\cdot)$ and $\grave{\boldsymbol{f}}(\cdot)$ to only contain orthogonal functions since one can always adjust the elements in $\boldsymbol{\varphi}_1(\cdot) \in \mathbb{L}^2([-r_1, 0] ; \mathbb{R}^{\mu_1})$ and $\boldsymbol{\varphi}_2(\cdot) \in \mathbb{L}^2([-r_2, -r_1] ; \mathbb{R}^{\mu_2})$ to satisfy (7). Note that $\boldsymbol{\varphi}_1(\cdot)$ and $\boldsymbol{\varphi}_2(\cdot)$ can become a $0\times 1$ empty vector if $\mu_1 = \mu_2 = 0$. Finally, the matrix inequalities in (8) can be verified via numerical calculations[2] with given $\acute{\boldsymbol{f}}(\cdot)$, $\grave{\boldsymbol{f}}(\cdot)$ and $\boldsymbol{\varphi}_1(\cdot)$, $\boldsymbol{\varphi}_2(\cdot)$.

**Remark 3.** (7) is employed in this paper to handle the distributed delay terms in (1) so that a well-posed dissipativity and stability condition can be derived later. This will be illustrated later in light of the results in Lemma 3 and Theorem 1. It is worthy to stress that (1) generalizes all the models in considered in Fridman & Tsodik (2009); Seuret *et al.* (2015); Feng & Nguang (2016b) without considering uncertainties.

**Remark 4.** A neutral delay system

$$\frac{\mathsf{d}}{\mathsf{d}t}\left(\boldsymbol{y}(t) - A_4\boldsymbol{y}(t-r)\right) = A_1\boldsymbol{y}(t) + A_2\boldsymbol{y}(t-r) + \int_{-r}^{0} A_3(\tau)\boldsymbol{x}(t+\tau)\mathsf{d}\tau$$

can be equivalently expressed by a CDDS:

$$\begin{aligned}
\dot{\boldsymbol{x}}(t) &= A_1\boldsymbol{x}(t) + (A_2 + A_1A_4)\boldsymbol{y}(t-r) + \int_{-r}^{0} A_3(\tau)\boldsymbol{x}(t+\tau)\mathsf{d}\tau\\
\boldsymbol{y}(t) &= \boldsymbol{x}(t) + A_4\boldsymbol{y}(t-r).
\end{aligned}$$

On the other hand, if there is rank redundancy in the delay matrices, namely,

$$\frac{\mathsf{d}}{\mathsf{d}t}\left(\boldsymbol{y}(t) - A_4N\boldsymbol{y}(t-r)\right) = A_1\boldsymbol{y}(t) + A_2N\boldsymbol{y}(t-r) + \int_{-r}^{0} A_3(\tau)N\boldsymbol{y}(t+\tau)\mathsf{d}\tau, \tag{9}$$

then one can first change (9) into

$$\frac{\mathsf{d}}{\mathsf{d}t}\left(\boldsymbol{y}(t) - A_4\boldsymbol{z}(t-r)\right) = A_1\boldsymbol{y}(t) + A_2\boldsymbol{z}(t-r) + \int_{-r}^{0} A_3(\tau)\boldsymbol{z}(t+\tau)\mathsf{d}\tau, \quad \boldsymbol{z}(t) = N\boldsymbol{y}(t). \tag{10}$$

Furthermore, let $\boldsymbol{x}(t) = \boldsymbol{y}(t) - A_4\boldsymbol{z}(t-r)$ considering (10), one can obtain the equivalent CDDS representation

$$\dot{\boldsymbol{x}}(t) = A_1\boldsymbol{x}(t) + (A_1A_4 + A_2)\,\boldsymbol{z}(t-r) + \int_{-r}^{0} A_3(\tau)\boldsymbol{z}(t+\tau)\mathsf{d}\tau$$

[2]One option is to use `vpaintegral` in Matlab which performs high-precision numerical integration.

$$z(t) = Nx(t) + NA_4 z(t-r)$$

which now is clearly advantageous in terms of reducing the scale of dimensionality if $\dim[z(t)] \ll \dim[y(t)]$. Finally, for the exploitation the rank redundancies among the state space variables of the retarded cases, see Gu & Liu (2009) for details.

In this paper, the functions $\acute{\boldsymbol{f}}(\cdot)$ and $\grave{\boldsymbol{f}}(\cdot)$ in (7) are applied to approximate the functions $\boldsymbol{\varphi}_1(\cdot) \in \mathbb{L}^2([-r_1, 0] ; \mathbb{R}^{\mu_1})$ and $\boldsymbol{\varphi}_2(\cdot) \in \mathbb{L}^2([-r_2, -r_1] ; \mathbb{R}^{\mu_2})$ in (7), respectively, where $\boldsymbol{\varphi}_1(\cdot)$ and $\boldsymbol{\varphi}_2(\cdot)$ might not satisfy (3). Specifically, the approximations are denoted by the decomposition:

$$\boldsymbol{\varphi}_1(\tau) = \acute{\Gamma}_d \acute{\boldsymbol{f}}(\tau) + \acute{\boldsymbol{\varepsilon}}_d(\tau), \quad \boldsymbol{\varphi}_2(\tau) = \grave{\Gamma}_\delta \grave{\boldsymbol{f}}(\tau) + \grave{\boldsymbol{\varepsilon}}_\delta(\tau) \tag{11}$$

where $\acute{\Gamma}_d$ and $\grave{\Gamma}_\delta$ are given coefficient. Furthermore, $\acute{\boldsymbol{\varepsilon}}_d(\tau) = \boldsymbol{\varphi}_1(\tau) - \acute{\Gamma}_d \acute{\boldsymbol{f}}(\tau)$ and $\grave{\boldsymbol{\varepsilon}}_\delta(\tau) = \boldsymbol{\varphi}_2(\tau) - \grave{\Gamma}_\delta \grave{\boldsymbol{f}}(\tau)$ contain the errors of approximations. In addition, we define matrices

$$\mathbb{S}^{\mu_1 \times \mu_1} \ni \acute{\mathsf{E}}_d := \int_{-r_1}^{0} \acute{\boldsymbol{\varepsilon}}_d(\tau) \acute{\boldsymbol{\varepsilon}}_d^\top(\tau) \mathsf{d}\tau, \quad \mathbb{S}^{\mu_2 \times \mu_2} \ni \grave{\mathsf{E}}_\delta := \int_{-r_2}^{-r_1} \grave{\boldsymbol{\varepsilon}}_\delta(\tau) \grave{\boldsymbol{\varepsilon}}_\delta^\top(\tau) \mathsf{d}\tau \tag{12}$$

to measure the error residues of (11). Inspired by the idea of orthogonal approximation in Hilbert space Muscat (2014), one option for the values of $\acute{\Gamma}_d$ and $\grave{\Gamma}_\delta$ in (11) is

$$\begin{aligned} \mathbb{R}^{\mu_1 \times d} \ni \acute{\Gamma}_d &:= \int_{-r_1}^{0} \boldsymbol{\varphi}_1(\tau) \acute{\boldsymbol{f}}^\top(\tau) \mathsf{d}\tau \acute{\mathsf{F}}_d, \ \acute{\mathsf{F}}_d^{-1} = \int_{-r_2}^{-r_1} \acute{\boldsymbol{f}}(\tau) \acute{\boldsymbol{f}}^\top(\tau) \mathsf{d}\tau \\ \mathbb{R}^{\mu_2 \times \delta} \ni \grave{\Gamma}_\delta &:= \int_{-r_2}^{-r_1} \boldsymbol{\varphi}_2(\tau) \grave{\boldsymbol{f}}^\top(\tau) \mathsf{d}\tau \acute{\mathsf{F}}_\delta, \ \grave{\mathsf{F}}_\delta^{-1} = \int_{-r_2}^{-r_1} \grave{\boldsymbol{f}}(\tau) \grave{\boldsymbol{f}}^\top(\tau) \mathsf{d}\tau. \end{aligned} \tag{13}$$

**Remark 5.** (13) might be interpreted as a vector form of the standard approximations (Least Squares) in Hilbert space. (See section 10.2 in Muscat (2014)) If $\acute{\boldsymbol{f}}(\cdot)$ and $\grave{\boldsymbol{f}}(\cdot)$ in (13) contains only Legendre polynomials, then (11)–(13) generalizes the polynomials approximation scheme proposed in Seuret *et al.* (2015) via a matrix framework. Finally, it is very crucial to emphasize that (11) does not restrict one only to apply (13) for the values of $\acute{\Gamma}_d$ and $\grave{\Gamma}_\delta$. Other appropriate options for $\acute{\Gamma}_d$ and $\grave{\Gamma}_\delta$ can be considered as well based on specific contexts.

The following property of Kronecker products will be used throughout the paper.

**Lemma 1.** *$\forall X \in \mathbb{R}^{n \times m}$, $\forall Y \in \mathbb{R}^{m \times p}$, $\forall Z \in \mathbb{R}^{q \times r}$,*

$$(X \otimes I_q)(Y \otimes Z) = (XY) \otimes (I_q Z) = (XY) \otimes Z = (XY) \otimes (Z I_r) = (X \otimes Z)(Y \otimes I_r). \tag{14}$$

*Moreover, $\forall X \in \mathbb{R}^{n \times m}$, we have*

$$\begin{bmatrix} A & B \\ C & D \end{bmatrix} \otimes X = \begin{bmatrix} A \otimes X & B \otimes X \\ C \otimes X & D \otimes X \end{bmatrix} \tag{15}$$

*for any $A, B, C, D$ with appropriate dimensions.*

The system (1) can be re-expressed as

$$\begin{aligned} &\dot{\boldsymbol{x}}(t) = \mathbf{A}\boldsymbol{\vartheta}(t), \quad \boldsymbol{y}(t) = \begin{bmatrix} \mathsf{O}_{\nu \times (2\nu+q)} & \Xi & \mathsf{O}_{\nu \times \nu\mu} \end{bmatrix} \boldsymbol{\vartheta}(t), \quad \boldsymbol{z}(t) = \Sigma \boldsymbol{\vartheta}(t) \\ &\boldsymbol{x}(t_0) = \boldsymbol{\xi} \in \mathbb{R}^n, \quad \forall \theta \in [-r_2, 0], \ \boldsymbol{y}(t_0 + \theta) = \boldsymbol{\psi}(\theta) \end{aligned} \tag{16}$$

with

$$\begin{aligned} \mathbf{A} = \Bigg[ \mathsf{O}_{n \times 2\nu} \quad D_1 \quad A_1 \quad A_2 \quad A_3 \quad A_4 \left( \begin{bmatrix} \acute{\Gamma}_d \\ I_d \end{bmatrix} \otimes I_\nu \right) \quad A_5 \left( \begin{bmatrix} \grave{\Gamma}_\delta \\ I_\delta \end{bmatrix} \otimes I_\nu \right) \cdots \\ \cdots A_4 \left( \begin{bmatrix} \acute{\mathsf{E}}_d \\ \mathsf{O}_{d \times \mu_1} \end{bmatrix} \otimes I_\nu \right) \quad A_5 \left( \begin{bmatrix} \grave{\mathsf{E}}_\delta \\ \mathsf{O}_{\delta \times \mu_2} \end{bmatrix} \otimes I_\nu \right) \Bigg] \end{aligned} \tag{17}$$

$$\Xi = \begin{bmatrix} A_6 & A_7 & A_8 & \mathsf{O}_{\nu\times\varrho\nu}\end{bmatrix} \tag{18}$$

$$\begin{aligned}\Sigma = \Bigg[ C_6 \quad C_7 \quad D_2 \quad C_1 \quad C_2 \quad C_3 \quad C_4\left(\begin{bmatrix}\acute{\Gamma}_d\\ I_d\end{bmatrix}\otimes I_\nu\right) \quad C_5\left(\begin{bmatrix}\grave{\Gamma}_\delta\\ I_\delta\end{bmatrix}\otimes I_\nu\right)\cdots \\ \cdots C_4\left(\begin{bmatrix}\acute{\mathsf{E}}_d\\ \mathsf{O}_{d\times\mu_1}\end{bmatrix}\otimes I_\nu\right) \quad C_5\left(\begin{bmatrix}\grave{\mathsf{E}}_\delta\\ \mathsf{O}_{\delta\times\mu_2}\end{bmatrix}\otimes I_\nu\right)\Bigg]\end{aligned} \tag{19}$$

$$\boldsymbol{\vartheta}(t) := \mathbf{Col}\left(\begin{bmatrix}\dot{\boldsymbol{y}}(t-r_1)\\ \dot{\boldsymbol{y}}(t-r_2)\end{bmatrix}, \begin{bmatrix}\boldsymbol{w}(t)\\ \boldsymbol{x}(t)\end{bmatrix}, \begin{bmatrix}\boldsymbol{y}(t-r_1)\\ \boldsymbol{y}(t-r_2)\end{bmatrix}, \begin{bmatrix}\int_{-r_1}^{0}\acute{F}_d(\tau)\boldsymbol{y}(t+\tau)\mathsf{d}\tau\\ \int_{-r_2}^{-r_1}\grave{F}_\delta(\tau)\boldsymbol{y}(t+\tau)\mathsf{d}\tau\end{bmatrix}, \begin{bmatrix}\int_{-r_1}^{0}\acute{E}_d(\tau)\boldsymbol{y}(t+\tau)\mathsf{d}\tau\\ \int_{-r_2}^{-r_1}\grave{E}_\delta(\tau)\boldsymbol{y}(t+\tau)\mathsf{d}\tau\end{bmatrix}\right), \tag{20}$$

where $\mathbb{R}^{d\nu\times\nu} \ni \acute{F}_d(\tau) := \acute{\boldsymbol{f}}(\tau)\otimes I_\nu$ and $\mathbb{R}^{\delta\nu\times\nu} \ni \grave{F}_\delta(\tau) := \grave{\boldsymbol{f}}(\tau)\otimes I_\nu$ and $\acute{E}_d(\tau) := \acute{\mathsf{E}}_d^{-1}\acute{\boldsymbol{\varepsilon}}_d(\tau)\otimes I_\nu$ and $\grave{E}_\delta(\tau) := \grave{\mathsf{E}}_\delta^{-1}\grave{\boldsymbol{\varepsilon}}_\delta(\tau)\otimes I_\nu$ with $\acute{\mathsf{E}}_d$ and $\grave{\mathsf{E}}_\delta$ in (12). Note that $\acute{\mathsf{E}}_d$ and $\grave{\mathsf{E}}_\delta$ in (12) are invertible according to what will be explained in Remark 9 based on what will be presented in (26) and (29). Note that also the distributed delay terms in (16) are derived based on the identities

$$\begin{aligned}&\left(\begin{bmatrix}\boldsymbol{\varphi}_1(\tau)\\ \acute{\boldsymbol{f}}(\tau)\end{bmatrix}\otimes I_\nu\right)\boldsymbol{y}(t+\tau) = \left(\begin{bmatrix}\acute{\Gamma}_d\acute{\boldsymbol{f}}(\tau)+\acute{\boldsymbol{\varepsilon}}_d(\tau)\\ \acute{\boldsymbol{f}}(\tau)\end{bmatrix}\otimes I_\nu\right)\boldsymbol{y}(t+\tau) = \left(\begin{bmatrix}\acute{\Gamma}_d\\ I_d\end{bmatrix}\acute{\boldsymbol{f}}(\tau)\otimes I_\nu\right)\boldsymbol{y}(t+\tau)\\ &\quad + \left(\begin{bmatrix}I_{\mu_1}\\ \mathsf{O}_{d\times\mu_1}\end{bmatrix}\acute{\boldsymbol{\varepsilon}}_d(\tau)\otimes I_\nu\right)\boldsymbol{y}(t+\tau) = \left(\begin{bmatrix}\acute{\Gamma}_d\\ I_d\end{bmatrix}\otimes I_\nu\right)\acute{F}_d(\tau)\boldsymbol{y}(t+\tau) + \left(\begin{bmatrix}\acute{\mathsf{E}}_d\\ \mathsf{O}_{d\times\mu_1}\end{bmatrix}\otimes I_\nu\right)\acute{E}_d(\tau)\boldsymbol{y}(t+\tau)\\ &\left(\begin{bmatrix}\boldsymbol{\varphi}_2(\tau)\\ \grave{\boldsymbol{f}}(\tau)\end{bmatrix}\otimes I_\nu\right)\boldsymbol{y}(t+\tau) = \left(\begin{bmatrix}\grave{\Gamma}_\delta\grave{\boldsymbol{f}}(\tau)+\grave{\boldsymbol{\varepsilon}}_\delta(\tau)\\ \grave{\boldsymbol{f}}(\tau)\end{bmatrix}\otimes I_\nu\right)\boldsymbol{y}(t+\tau) = \left(\begin{bmatrix}\grave{\Gamma}_\delta\\ I_d\end{bmatrix}\grave{\boldsymbol{f}}(\tau)\otimes I_\nu\right)\boldsymbol{y}(t+\tau)\\ &\quad + \left(\begin{bmatrix}I_{\mu_2}\\ \mathsf{O}_{\delta\times\mu_2}\end{bmatrix}\grave{\boldsymbol{\varepsilon}}_\delta(\tau)\otimes I_\nu\right)\boldsymbol{y}(t+\tau) = \left(\begin{bmatrix}\grave{\Gamma}_\delta\\ I_\delta\end{bmatrix}\otimes I_\nu\right)\grave{F}_\delta(\tau)\boldsymbol{y}(t+\tau) + \left(\begin{bmatrix}\grave{\mathsf{E}}_\delta\\ \mathsf{O}_{\delta\times\mu_2}\end{bmatrix}\otimes I_\nu\right)\grave{E}_\delta(\tau)\boldsymbol{y}(t+\tau)\end{aligned}$$

which themselves are obtained via the property of Kronecker product in (14).

## 3. Mathematical preliminaries

In this section some important lemmas and definition are present. This includes a novel integral inequality which will be applied later for the derivation of our dissipative stability condition.

The following lemma provides sufficient conditions for the stability of (1). It can be interpreted as a particular case of Theorem 3 in Gu & Liu (2009) with certain modifications.

**Lemma 2.** *Given $r_2 > r_1 > 0$, the system (1) with $\boldsymbol{w}(t) \equiv \mathbf{0}_q$ is globally uniformly asymptotically stable at its origin if there exist $\epsilon_1; \epsilon_2; \epsilon_3 > 0$ and a differentiable functional $v : \mathbb{R}^n \times \mathcal{A}([-r_2, 0) ⨾ \mathbb{R}^\nu) \to \mathbb{T}$ such that $v(\mathbf{0}_n, \mathbf{0}_\nu) = 0$ and*

$$\epsilon_1\|\boldsymbol{\xi}\|_2^2 \le v(\boldsymbol{\xi}, \boldsymbol{\psi}(\cdot)) \le \epsilon_2\left[\|\boldsymbol{\xi}\|_2 \vee \left(\|\boldsymbol{\psi}(\cdot)\|_\infty + \|\dot{\boldsymbol{\psi}}(\cdot)\|_2\right)\right]^2, \tag{21}$$

$$\dot{v}(r, \boldsymbol{\xi}, \boldsymbol{\psi}(\cdot)) := \left.\frac{\mathsf{d}^+}{\mathsf{d}t}v(\boldsymbol{x}(t), \mathbf{y}_t(\cdot))\right|_{t=t_0, \boldsymbol{x}(t_0)=\boldsymbol{\xi}, \mathbf{y}_{t_0}(\cdot)=\boldsymbol{\psi}(\cdot)} \le -\epsilon_3\|\boldsymbol{\xi}\|_2^2 \tag{22}$$

*for any $\boldsymbol{\xi} \in \mathbb{R}^n$ and $\boldsymbol{\psi}(\cdot) \in \mathcal{A}([-r_2, 0) ⨾ \mathbb{R}^\nu)$ in (1), where $t_0 \in \mathbb{R}$ and $\frac{\mathsf{d}^+}{\mathsf{d}x}f(x) = \operatorname{limsup}_{\eta\downarrow 0}\frac{f(x+\eta)-f(x)}{\eta}$. Furthermore, $\mathbf{y}_t(\cdot)$ in (22) is defined by the equality $\forall t \ge t_0$, $\forall\theta \in [-r, 0)$, $\mathbf{y}_t(\theta) = \boldsymbol{y}(t+\theta)$ where $\boldsymbol{x}(t)$ and $\boldsymbol{y}(t)$ here satisfying (1) with $\boldsymbol{w}(t) \equiv \mathbf{0}_q$.*

**Definition 1 (Dissipativity).** Given $r_2 > r_1 > 0$, the coupled differential functional system (1) with a supply rate function $s(\boldsymbol{z}(t), \boldsymbol{w}(t))$ is said to be dissipative if there exists a differentiable functional $v : \mathbb{R}^n \times \mathcal{A}([-r_2, 0) ⨾ \mathbb{R}^\nu) \to \mathbb{R}$ such that

$$\forall t \ge t_0 : \ \dot{v}(\boldsymbol{x}(t), \mathbf{y}_t(\cdot)) - s(\boldsymbol{z}(t), \boldsymbol{w}(t)) \le 0 \tag{23}$$

where $\mathbf{y}_t(\cdot)$ is defined by the equality $\forall t \ge t_0$, $\forall\theta \in [-r_2, 0)$, $\mathbf{y}_t(\theta) = \boldsymbol{y}(t+\theta)$, and $\boldsymbol{x}(t)$, $\boldsymbol{y}(t)$ and $\boldsymbol{z}(t)$ satisfy the equialities in (1) with $\boldsymbol{w}(\cdot) \in \widehat{\mathbb{L}}^2([t_0, \infty) ⨾ \mathbb{R}^q)$.

In this paper, we apply the quadratic form

$$s(\boldsymbol{z}(t),\boldsymbol{w}(t)) = [*]\,\mathbf{J}\begin{bmatrix}\boldsymbol{z}(t)\\ \boldsymbol{w}(t)\end{bmatrix},\quad \mathbf{J}=\begin{bmatrix}J_1 & J_2\\ * & J_3\end{bmatrix},\quad \mathbb{S}^m \ni \widetilde{J}^\top J_1^{-1}\widetilde{J}\preceq 0,\ \ J_1^{-1}\prec 0,\ \ \widetilde{J}\in\mathbb{R}^{m\times m},\ \ J_3\in\mathbb{S}^q \tag{24}$$

as the supply function to characterize dissiaptivity. The form of $\mathbf{J}$ in (24) is constructed considering the general quadratic constraints in Scherer *et al.* (1997) together with the idea of the fatorization of the matrix $U_j$ in Scherer *et al.* (1997). Note that (24) is able to characterize numerous performance criteria such as

- $\mathbb{L}^2$ gain performance: $J_1 = -\gamma I_m$, $\widetilde{J} = I_m$, $J_2 = \mathsf{O}_{m\times q}$, $J_3 = \gamma I_q$ where $\gamma > 0$.
- Passivity: $J_1 \prec 0$, $\widetilde{J} = \mathsf{O}_m$, $J_2 = I_m$, $J_3 = \mathsf{O}_m$ with $m = q$.

*3.1. A novel integral inequality*

The following generalized new integral inequality is proposed which will be employed for the derivation of our major results on the dissipativity and stability analysis in this section. Firstly, we define the weighted Lebesgue function space

$$\mathbb{L}^2_{\varpi}\left(\mathcal{K}\,⨾\,\mathbb{R}^d\right) := \left\{\boldsymbol{\psi}(\cdot)\in\mathbb{L}_f\left(\mathcal{K}\,⨾\,\mathbb{R}^d\right) : \|\boldsymbol{\psi}(\cdot)\|_{2,\varpi} < \infty\right\} \tag{25}$$

with $d\in\mathbb{N}$ and $\|\boldsymbol{\psi}(\cdot)\|_{2,\varpi} := \int_{\mathcal{K}}\varpi(\tau)\boldsymbol{\psi}^\top(\tau)\boldsymbol{\psi}(\tau)\mathsf{d}\tau$, where $\varpi(\cdot)\in\mathbb{L}_f(\mathcal{K}\,⨾\,\mathbb{T})$ and the function $\varpi(\cdot)$ has only countable or finite number of zero values. Furthermore, $\mathcal{K}\subseteq\mathbb{R}\cup\{\pm\infty\}$ and $\int_{\mathcal{K}}\mathsf{d}\tau\neq 0$.

**Lemma 3.** *Given $\mathcal{K}$ and $\varpi(\cdot)$ in (25) and $U\in\mathbb{S}^n_{\succeq 0} := \{X\in\mathbb{S}^n : X\succeq 0\}$ with $n\in\mathbb{N}$. Let $\mathsf{f}(\cdot) := \mathbf{Col}_{i=1}^d\,\mathsf{f}_i(\cdot)\in\mathbb{L}^2_{\varpi}\left(\mathcal{K}\,⨾\,\mathbb{R}^d\right)$ and $\mathsf{g}(\cdot) := \mathbf{Col}_{i=1}^\delta\,\mathsf{g}_i(\cdot)\in\mathbb{L}^2_{\varpi}\left(\mathcal{K}\,⨾\,\mathbb{R}^\delta\right)$ with $d\in\mathbb{N}$ and $\delta\in\mathbb{N}_0$, in which the functions $\mathsf{f}(\cdot)$ and $\mathsf{g}(\cdot)$ satisfy*

$$\int_{\mathcal{K}}\varpi(\tau)\begin{bmatrix}\mathsf{g}(\tau)\\ \mathsf{f}(\tau)\end{bmatrix}\begin{bmatrix}\mathsf{g}^\top(\tau) & \mathsf{f}^\top(\tau)\end{bmatrix}\mathsf{d}\tau\succ 0. \tag{26}$$

*Then we have,*

$$\begin{aligned}\forall\boldsymbol{x}(\cdot)\in\mathbb{L}^2_{\varpi}(\mathcal{K}\,⨾\,\mathbb{R}^n),\ \int_{\mathcal{K}}\varpi(\tau)\boldsymbol{x}^\top(\tau)U\boldsymbol{x}(\tau)\mathsf{d}\tau &\geq \int_{\mathcal{K}}\varpi(\tau)\boldsymbol{x}^\top(\tau)\mathsf{F}^\top(\tau)\mathsf{d}\tau\left(\mathcal{F}_d\otimes U\right)\int_{\mathcal{K}}\varpi(\tau)\mathsf{F}(\tau)\boldsymbol{x}(\tau)\mathsf{d}\tau\\ &+\int_{\mathcal{K}}\varpi(\tau)\boldsymbol{x}^\top(\tau)\mathsf{E}^\top(\tau)\mathsf{d}\tau\left(\mathcal{E}_d^{-1}\otimes U\right)\int_{\mathcal{K}}\varpi(\tau)\mathsf{E}(\tau)\boldsymbol{x}(\tau)\mathsf{d}\tau\end{aligned} \tag{27}$$

*where*

$$\begin{aligned}&\mathsf{F}(\tau) = \mathsf{f}(\tau)\otimes I_n\in\mathbb{R}^{dn\times n},\quad \mathcal{F}_d^{-1} = \int_{\mathcal{K}}\varpi(\tau)\mathsf{f}(\tau)\mathsf{f}^\top(\tau)\mathsf{d}\tau\in\mathbb{S}^d_{\succ 0}\\ &\mathsf{E}(\tau) = \boldsymbol{\mathsf{e}}(\tau)\otimes I_n\in\mathbb{R}^{\delta n\times n},\quad \mathcal{E}_d = \int_{\mathcal{K}}\varpi(\tau)\boldsymbol{\mathsf{e}}(\tau)\boldsymbol{\mathsf{e}}^\top(\tau)\mathsf{d}\tau\in\mathbb{S}^\delta\\ &\boldsymbol{\mathsf{e}}(\tau) = \mathsf{g}(\tau)-\mathsf{A}\mathsf{f}(\tau)\in\mathbb{R}^\delta,\quad \mathsf{A} = \int_{\mathcal{K}}\varpi(\tau)\mathsf{g}(\tau)\mathsf{f}^\top(\tau)\mathsf{d}\tau\mathcal{F}_d\in\mathbb{R}^{\delta\times d}.\end{aligned} \tag{28}$$

*Proof.* The proof of Lemma 3 is inspired by the proof of Lemma 2 in Seuret *et al.* (2015) and the proof of Lemma 5 in Feng & Nguang (2016b). Firstly, one can conclude that $\mathcal{E}_d$ in (28) is invertible for any $\mathsf{f}(\cdot)\in\mathbb{L}^2_{\varpi}\left(\mathcal{K}\,⨾\,\mathbb{R}^d\right)$, $\mathsf{g}(\cdot)\in\mathbb{L}^2_{\varpi}\left(\mathcal{K}\,⨾\,\mathbb{R}^\delta\right)$ satisfying (26) since

$$\mathcal{E}_d = \int_{\mathcal{K}}\varpi(\tau)\boldsymbol{\mathsf{e}}(\tau)\boldsymbol{\mathsf{e}}^\top(\tau)\mathsf{d}\tau = \begin{bmatrix}I_\delta & -\mathsf{A}\end{bmatrix}\int_{\mathcal{K}}\varpi(\tau)\begin{bmatrix}\mathsf{g}(\tau)\\ \mathsf{f}(\tau)\end{bmatrix}\begin{bmatrix}\mathsf{g}^\top(\tau) & \mathsf{f}^\top(\tau)\end{bmatrix}\mathsf{d}\tau\begin{bmatrix}I_\delta & -\mathsf{A}\end{bmatrix}^\top\succ 0, \tag{29}$$

where the positive definite matrix inequality can be derived based on (26) and the property of congruence transformations with the fact that rank $\begin{bmatrix}I_\delta & -\mathsf{A}\end{bmatrix} = \delta$. Consequently, $\mathcal{E}_d^{-1}$ is well defined.

Let $\boldsymbol{v}(\tau) := \boldsymbol{x}(\tau) - \mathsf{F}^\top(\tau)(\mathcal{F}_d \otimes I_n) \int_{\mathcal{K}} \varpi(\theta)\mathsf{F}(\theta)\boldsymbol{x}(\theta)\mathsf{d}\theta - \mathsf{E}^\top(\tau) \left(\mathcal{E}_d^{-1} \otimes I_n\right) \int_{\mathcal{K}} \varpi(\theta)\mathsf{E}(\theta)\boldsymbol{x}(\theta)\mathsf{d}\theta$, where $\mathsf{F}(\cdot)$, $\mathsf{E}(\cdot)$ have been given in Lemma 3. By $\mathsf{A} = \int_{\mathcal{K}} \varpi(\tau)\mathbf{g}(\tau)\mathbf{f}^\top(\tau)\mathsf{d}\tau\mathcal{F}_d$ and $\boldsymbol{e}(\tau) = \mathbf{g}(\tau) - \mathsf{A}\mathbf{f}(\tau) \in \mathbb{R}^\delta$, we have

$$\begin{aligned}\int_{\mathcal{K}} \varpi(\tau)\boldsymbol{e}(\tau)\mathbf{f}^\top(\tau)\mathsf{d}\tau &= \int_{\mathcal{K}} \varpi(\tau)\left[\mathbf{g}(\tau) - \mathsf{A}\mathbf{f}(\tau)\right]\mathbf{f}^\top(\tau)\mathsf{d}\tau = \int_{\mathcal{K}} \varpi(\tau)\mathbf{g}(\tau)\mathbf{f}^\top(\tau)\mathsf{d}\tau - \mathsf{A}\int_{\mathcal{K}} \varpi(\tau)\mathbf{f}(\tau)\mathbf{f}^\top(\tau)\mathsf{d}\tau \\ &= \int_{\mathcal{K}} \varpi(\tau)\mathbf{g}(\tau)\mathbf{f}^\top(\tau)\mathsf{d}\tau - \left(\int_{\mathcal{K}} \varpi(\tau)\mathbf{g}(\tau)\mathbf{f}^\top(\tau)\mathsf{d}\tau\right)\mathcal{F}_d\mathcal{F}_d^{-1} = \mathsf{O}_{\delta\times d}. \quad (30)\end{aligned}$$

Now substituting the expression of $\boldsymbol{v}(\cdot)$ into $\int_{\mathcal{K}} \varpi(\tau)\boldsymbol{v}^\top(\tau)U\boldsymbol{v}(\tau)\mathsf{d}\tau$ and considering (30) yields

$$\begin{aligned}\int_{\mathcal{K}} \varpi(\tau)\boldsymbol{v}^\top(\tau)U\boldsymbol{v}(\tau)\mathsf{d}\tau &= \int_{\mathcal{K}} \varpi(\tau)\boldsymbol{x}^\top(\tau)U\boldsymbol{x}(\tau)\mathsf{d}\tau - 2\int_{\mathcal{K}} \varpi(\tau)\boldsymbol{x}^\top(\tau)U\mathsf{F}^\top(\tau)\mathsf{d}\tau(\mathcal{F}_d \otimes I_n)\boldsymbol{\zeta} \\ &+ \boldsymbol{\zeta}^\top\int_{\mathcal{K}} \varpi(\tau)(\mathcal{F}_d \otimes I_n)^\top\mathsf{F}(\tau)U\mathsf{F}^\top(\tau)(\mathcal{F}_d \otimes I_n)\mathsf{d}\tau\boldsymbol{\zeta} - 2\int_{\mathcal{K}} \varpi(\tau)\boldsymbol{x}^\top(\tau)U\mathsf{E}^\top(\tau)\mathsf{d}\tau(\mathcal{E}_d^{-1} \otimes I_n)\boldsymbol{\omega} \\ &+ \boldsymbol{\omega}^\top\int_{\mathcal{K}} \varpi(\tau)(\mathcal{E}_d^{-1} \otimes I_n)^\top\mathsf{E}(\tau)U\mathsf{E}^\top(\tau)(\mathcal{E}_d^{-1} \otimes I_n)\mathsf{d}\tau\boldsymbol{\omega}\end{aligned} \quad (31)$$

where $\boldsymbol{\zeta} := \int_{\mathcal{K}} \varpi(\theta)\mathsf{F}(\theta)\boldsymbol{x}(\theta)\mathsf{d}\theta$ and $\boldsymbol{\omega} := \int_{\mathcal{K}} \varpi(\theta)\mathsf{E}(\theta)\boldsymbol{x}(\theta)\mathsf{d}\theta$. Apply (14) to the term $U\mathsf{F}^\top(\tau)$ and $U\mathsf{E}^\top(\tau)$ and consider $\mathsf{F}(\tau) = \mathbf{f}(\tau) \otimes I_n$ and $\mathsf{E}(\tau) = \boldsymbol{e}(\tau) \otimes I_n$, then we have

$$U\mathsf{F}^\top(\tau) = \mathsf{F}^\top(\tau)(I_d \otimes U), \quad U\mathsf{E}^\top(\tau) = \mathsf{E}^\top(\tau)(I_\delta \otimes U) \quad (32)$$

given $(X \otimes Y)^\top = X^\top \otimes Y^\top$. One the other hand, it is true that

$$\begin{aligned}\int_{\mathcal{K}} \varpi(\tau)\mathsf{F}(\tau)\mathsf{F}^\top(\tau)\mathsf{d}\tau &= \left(\int_{\mathcal{K}} \varpi(\tau)\mathbf{f}(\tau)\mathbf{f}^\top(\tau)\mathsf{d}\tau\right) \otimes I_n = \mathcal{F}_d^{-1} \otimes I_n \\ \int_{\mathcal{K}} \varpi(\tau)\mathsf{E}(\tau)\mathsf{E}^\top(\tau)\mathsf{d}\tau &= \left(\int_{\mathcal{K}} \varpi(\tau)\boldsymbol{e}(\tau)\boldsymbol{e}^\top(\tau)\mathsf{d}\tau\right) \otimes I_n = \mathcal{E}_d \otimes I_n\end{aligned} \quad (33)$$

since $\mathsf{F}(\tau) = \mathbf{f}(\tau) \otimes I_n$ and $\mathsf{E}(\tau) = \boldsymbol{e}(\tau) \otimes I_n$. By using (32) and (33) with (14) to some of the terms in (31), it follows that

$$\begin{aligned}\int_{\mathcal{K}} \varpi(\tau)\boldsymbol{x}^\top(\tau)U\mathsf{F}^\top(\tau)\mathsf{d}\tau(\mathcal{F}_d \otimes I_n)\boldsymbol{\zeta} &= \boldsymbol{\zeta}^\top(\mathcal{F}_d \otimes U)\boldsymbol{\zeta} \\ \int_{\mathcal{K}} \varpi(\tau)\boldsymbol{x}^\top(\tau)U\mathsf{E}^\top(\tau)\mathsf{d}\tau(\mathcal{E}_d^{-1} \otimes I_n)\boldsymbol{\omega} &= \boldsymbol{\omega}^\top(\mathcal{E}_d^{-1} \otimes U)\boldsymbol{\omega}.\end{aligned} \quad (34)$$

and

$$\begin{aligned}\int_{\mathcal{K}} (\mathcal{F}_d \otimes I_n)^\top\varpi(\tau)\mathsf{F}(\tau)U\mathsf{F}^\top(\tau)(\mathcal{F}_d \otimes I_n)\mathsf{d}\tau &= (\mathcal{F}_d \otimes I_n)\int_{\mathcal{K}} \varpi(\tau)\mathsf{F}(\tau)\mathsf{F}^\top(\tau)\mathsf{d}\tau(\mathcal{F}_d \otimes U) = \mathcal{F}_d \otimes U \\ \int_{\mathcal{K}} (\mathcal{E}_d^{-1} \otimes I_n)^\top\varpi(\tau)\mathsf{E}(\tau)U\mathsf{E}^\top(\tau)(\mathcal{E}_d^{-1} \otimes I_n)\mathsf{d}\tau &= (\mathcal{E}_d^{-1} \otimes I_n)\int_{\mathcal{K}} \varpi(\tau)\mathsf{E}(\tau)\mathsf{E}^\top(\tau)\mathsf{d}\tau\left(\mathcal{E}_d^{-1} \otimes U\right) \\ &= \mathcal{E}_d^{-1} \otimes U.\end{aligned} \quad (35)$$

Substituting (35) into (31) and also considering the relations in (34) yields

$$\begin{aligned}\int_{\mathcal{K}} \varpi(\tau)\boldsymbol{v}^\top(\tau)U\boldsymbol{v}(\tau)\mathsf{d}\tau =& \int_{\mathcal{K}} \varpi(\tau)\boldsymbol{x}^\top(\tau)U\boldsymbol{x}(\tau)\mathsf{d}\tau - \int_{\mathcal{K}} \varpi(\tau)\boldsymbol{x}^\top(\tau)\mathsf{F}^\top(\tau)\mathsf{d}\tau\,(\mathcal{F}_d \otimes U)\int_{\mathcal{K}} \varpi(\tau)\mathsf{F}(\tau)\boldsymbol{x}(\tau)\mathsf{d}\tau \\ &- \int_{\mathcal{K}} \varpi(\tau)\boldsymbol{x}^\top(\tau)\mathsf{E}^\top(\tau)\mathsf{d}\tau\left(\mathcal{E}_d^{-1} \otimes U\right)\int_{\mathcal{K}} \varpi(\tau)\mathsf{E}(\tau)\boldsymbol{x}(\tau)\mathsf{d}\tau. \quad (36)\end{aligned}$$

Given $U \succeq 0$, (36) gives (27). This finishes the proof. □

**Remark 6.** By Theorem 7.2.10 in Horn & Johnson (2012) and considering the fact that

$$\left(\mathbb{L}^2_{\varpi}\left(\mathcal{K} \,⨾\, \mathbb{R}^d\right) / \operatorname{Ker}\left(\|\bullet\|_{2,\varpi}\right), \quad \int_{\mathcal{K}} \varpi(\tau)\bullet_1^\top(\tau)\bullet_2(\tau)\mathsf{d}\tau\right)$$

is an inner product space[3], we know (26) indicates that the functions in $\mathsf{f}(\cdot)$ and $\mathsf{g}(\cdot)$ are linearly independent in a Lebesgue sense.

The following inequality can be obtained by setting $\delta = 0$ in Lemma 3 based on the notion of empty matrices.

**Corollary 1.** *Given $\mathcal{K}$ and $\varpi(\cdot)$ in (25) and $U \in \mathbb{S}^n_{\succeq 0} := \{X \in \mathbb{S}^n : X \succeq 0\}$ with $n \in \mathbb{N}$. Let $\mathsf{f}(\cdot) := \mathbf{Col}_{i=1}^d \mathsf{f}_i(\cdot) \in \mathbb{L}^2_{\varpi}\left(\mathcal{K} \,⨾\, \mathbb{R}^d\right)$ with $d \in \mathbb{N}$ where $\mathsf{f}(\cdot)$ satisfies*

$$\int_{\mathcal{K}} \varpi(\tau)\mathsf{f}(\tau)\mathsf{f}^\top(\tau)\mathsf{d}\tau \succ 0. \tag{37}$$

*Then the inequality*

$$\int_{\mathcal{K}} \varpi(\tau)\boldsymbol{x}^\top(\tau)U\boldsymbol{x}(\tau)\mathsf{d}\tau \geq \int_{\mathcal{K}} \varpi(\tau)\boldsymbol{x}^\top(\tau)\mathsf{F}^\top(\tau)\mathsf{d}\tau \left(\mathcal{F}_d \otimes U\right) \int_{\mathcal{K}} \varpi(\tau)\mathsf{F}(\tau)\boldsymbol{x}(\tau)\mathsf{d}\tau \tag{38}$$

*holds for all $\boldsymbol{x}(\cdot) \in \mathbb{L}^2_{\varpi}(\mathcal{K} \,⨾\, \mathbb{R}^n)$, where $\mathsf{F}(\tau) = \mathsf{f}(\tau) \otimes I_n \in \mathbb{R}^{dn \times n}$ and $\mathcal{F}_d^{-1} = \int_{\mathcal{K}} \varpi(\tau)\mathsf{f}(\tau)\mathsf{f}^\top(\tau)\mathsf{d}\tau \in \mathbb{S}^d_{\succ 0}$.*

**Remark 7.** (27) reduces to Lemma 1 in Seuret *et al.* (2015) if $\boldsymbol{f}(\cdot)$ contains only Legendre polynomials, that is, $\delta = 0$ and $\{f_i(\cdot)\}_{i=0}^d$ to be Legendre polynomials. Moreover, by utilizing the Cauchy formula for repeated integrations (see (5),(6) and (25),(26) in Gyurkovics & Takács (2016)), the results in Park *et al.* (2015); Gyurkovics & Takács (2016) and Chen *et al.* (2016) are covered by (38) as special cases with appropriate $\boldsymbol{f}(\cdot)$. Meanwhile, if $\boldsymbol{f}(\cdot)$ contains only orthogonal functions, then (38) reduces to the inequalities in Feng & Nguang (2016a) with a reverse order of Kronecker product. In addition, with $\mathcal{K} = [0 + \infty]$, (9) in Liu *et al.* (2016) is the special case of (38) with appropriate $\varpi(\cdot)$ and $\boldsymbol{f}(\cdot)$. By letting $\varpi(\tau) = 1$, (38) reduces to the result of Lemma 4 in Feng & Nguang (2016b). Finally, it is worthy to note that a summation inequality

$$\sum_{k\in\mathcal{J}} \varpi(k)[*]U\boldsymbol{x}(k) \geq [*]\left(\mathsf{F} \otimes U\right)\sum_{k\in\mathcal{J}} \varpi(k)F(k)\boldsymbol{x}(k), \quad \mathsf{F}^{-1} = \sum_{k\in\mathcal{J}} \varpi(k)\boldsymbol{f}(k)\boldsymbol{f}^\top(k) \quad \mathcal{J} \subseteq \mathbb{Z}, \ \#\mathcal{J} \geq 2 \tag{39}$$

with the prerequisite $\mathsf{F}^{-1} = \sum_{k\in\mathcal{J}} \varpi(k)\boldsymbol{f}(k)\boldsymbol{f}^\top(k) \succ 0$ can be easily obtained based on (38).Note that for a discrete system with finite length of delays indicating finite dimensions, (39) may produce a perfect bound with no conservatism at a finite $d$.

An interesting corollary of Lemma 3 is presented as follows which can be interpreted as a generalization of Lemma 1 in Seuret *et al.* (2015).

**Corollary 2.** *Given all the parameters defined in Lemma 3 with $\{\mathsf{f}_i(\tau)\}_{i=1}^\infty$ and $\mathsf{f}(\cdot) = \mathbf{Col}_{i=1}^d \mathsf{f}_i(\cdot)$ satisfying*

$$\forall d \in \mathbb{N}, \ \mathcal{F}_d^{-1} = \int_{\mathcal{K}} \mathsf{f}(\tau)\mathsf{f}^\top(\tau)\mathsf{d}\tau = \bigoplus_{i=1}^d \left(\int_{\mathcal{K}} \varpi(\tau)\mathsf{f}_i^2(\tau)\mathsf{d}\tau\right), \tag{40}$$

*then we have that*

$$\forall d \in \mathbb{N}, \ 0 \prec \mathcal{E}_{d+1} = \mathcal{E}_d - \left(\int_{\mathcal{K}} \varpi(\tau)\mathsf{f}_{d+1}^2(\tau)\mathsf{d}\tau\right)\mathsf{a}_{d+1}\mathsf{a}_{d+1}^\top \preceq \mathcal{E}_d \tag{41}$$

*where $\mathcal{E}_d$ is given in Lemma 3 and $\mathsf{a}_{d+1} := \left(\int_{\mathcal{K}} \varpi(\tau)\mathsf{g}(\tau)\mathsf{f}_{d+1}(\tau)\mathsf{d}\tau\right)\left(\int_{\mathcal{K}} \varpi(\tau)\mathsf{f}_{d+1}^2(\tau)\mathsf{d}\tau\right)^{-1} \in \mathbb{R}^\delta$ and $\mathsf{f}_{d+1}(\cdot) \in \mathbb{L}^2_{\varpi}(\mathcal{K} \,⨾\, \mathbb{R})$.*

[3] $\operatorname{Ker}\left(\|\bullet\|_{2,\varpi}\right) := \{\boldsymbol{f}(\cdot) \in \mathbb{L}^2_{\varpi}\left(\mathcal{K} \,⨾\, \mathbb{R}^d\right) : \|\boldsymbol{f}(\cdot)\|_{2,\varpi} = \mathbf{0}_d\}$

*Proof.* Note that only the dimension of $\mathbf{f}(\cdot)$ is related to $d$, whereas $\delta$ as the dimension of $\mathbf{g}(\cdot)$ is independent from $d$. It is obvious to see that given $\mathbf{f}(\cdot)$ satisfying (40), we have $\mathcal{F}_{d+1} = \mathcal{F}_d \oplus \left(\int_{\mathcal{K}} \varpi(\tau)\mathsf{f}_{d+1}^2(\tau)\mathrm{d}\tau\right)^{-1}$ (See the Definition 1 in Feng & Nguang (2016a)). By using this property, it follows that for all $d \in \mathbb{N}$

$$\boldsymbol{e}_{d+1}(\tau) = \mathbf{g}(\tau) - \left(\int_{\mathcal{K}} \varpi(\tau)\mathbf{g}(\tau)\begin{bmatrix}\mathbf{f}^\top(\tau) & \mathsf{f}_{d+1}(\tau)\end{bmatrix}\mathrm{d}\tau\right)\left[\mathcal{F}_d \oplus \left(\int_{\mathcal{K}} \varpi(\tau)\mathsf{f}_{d+1}^2(\tau)\mathrm{d}\tau\right)^{-1}\right]\begin{bmatrix}\mathbf{f}(\tau)\\ \mathsf{f}_{d+1}(\tau)\end{bmatrix} = \mathbf{g}(\tau) - \begin{bmatrix}\mathsf{A}_d & \boldsymbol{\alpha}_{d+1}\end{bmatrix}\begin{bmatrix}\mathbf{f}(\tau)\\ \mathsf{f}_{d+1}(\tau)\end{bmatrix} = \boldsymbol{e}_d(\tau) - \mathsf{f}_{d+1}(\tau)\boldsymbol{\alpha}_{d+1} \quad (42)$$

where $\boldsymbol{\alpha}_{d+1}$ has been defined in (41) and $\boldsymbol{e}_d(\tau) = \mathbf{g}(\tau) - \mathsf{A}_d\mathbf{f}(\tau)$. Note that the index $d$ is added to the symbols $\mathsf{A}$ and $\boldsymbol{e}(\tau)$ in Lemma 3 without causing ambiguity. By (42) and (29), we have

$$0 \prec \mathcal{E}_{d+1} = \int_{\mathcal{K}} \varpi(\tau)\boldsymbol{e}_{d+1}(\tau)\boldsymbol{e}_{d+1}^\top(\tau)\mathrm{d}\tau = \mathcal{E}_d - \mathbf{Sy}\left(\boldsymbol{\alpha}_{d+1}\int_{\mathcal{K}} \varpi(\tau)\mathsf{f}_{d+1}(\tau)\boldsymbol{e}_d^\top(\tau)\mathrm{d}\tau\right) + \left(\int_{\mathcal{K}} \varpi(\tau)\mathsf{f}_{d+1}^2(\tau)\mathrm{d}\tau\right)\boldsymbol{\alpha}_{d+1}\boldsymbol{\alpha}_{d+1}^\top. \quad (43)$$

By (30) and the fact that $\int_{\mathcal{K}} \varpi(\tau)\mathsf{f}_{d+1}(\tau)\mathbf{f}(\tau)\mathrm{d}\tau = \mathbf{0}_d$ due to (40), we have

$$\begin{aligned}\mathsf{O}_{\delta\times(d+1)} &= \int_{\mathcal{K}} \varpi(\tau)\boldsymbol{e}_{d+1}(\tau)\begin{bmatrix}\mathbf{f}^\top(\tau) & \mathsf{f}_{d+1}(\tau)\end{bmatrix}\mathrm{d}\tau = \int_{\mathcal{K}} \varpi(\tau)\left(\boldsymbol{e}_d(\tau) - \boldsymbol{\alpha}_{d+1}\mathsf{f}_{d+1}(\tau)\right)\begin{bmatrix}\mathbf{f}^\top(\tau) & \mathsf{f}_{d+1}(\tau)\end{bmatrix}\mathrm{d}\tau\\ &= \int_{\mathcal{K}} \varpi(\tau)\begin{bmatrix}\boldsymbol{e}_d(\tau)\mathbf{f}^\top(\tau) & \mathsf{f}_{d+1}(\tau)\boldsymbol{e}_d(\tau)\end{bmatrix}\mathrm{d}\tau - \boldsymbol{\alpha}_{d+1}\int_{\mathcal{K}} \varpi(\tau)\begin{bmatrix}\mathsf{f}_{d+1}(\tau)\mathbf{f}^\top(\tau) & \mathsf{f}_{d+1}^2(\tau)\end{bmatrix}\mathrm{d}\tau\\ &= \begin{bmatrix}\mathsf{O}_{\delta\times d} & \int_{\mathcal{K}} \varpi(\tau)\mathsf{f}_{d+1}(\tau)\boldsymbol{e}_d(\tau)\mathrm{d}\tau\end{bmatrix} - \begin{bmatrix}\mathsf{O}_{\delta\times d} & \int_{\mathcal{K}} \varpi(\tau)\mathsf{f}_{d+1}^2(\tau)\mathrm{d}\tau\boldsymbol{\alpha}_{d+1}\end{bmatrix} = \mathsf{O}_{\delta\times(d+1)}. \quad (44)\end{aligned}$$

Now (44) leads to the equality $\int_{\mathcal{K}} \varpi(\tau)\mathsf{f}_{d+1}(\tau)\boldsymbol{e}_d(\tau)\mathrm{d}\tau = \int_{\mathcal{K}} \varpi(\tau)\mathsf{f}_{d+1}^2(\tau)\mathrm{d}\tau\boldsymbol{\alpha}_{d+1}$. Substituting this equality into (43) yields (41) given $\int_{\mathcal{K}} \varpi(\tau)\mathsf{f}_{d+1}^2(\tau)\mathrm{d}\tau > 0$ and $\boldsymbol{\alpha}_{d+1}\boldsymbol{\alpha}_{d+1}^\top \succeq 0$. ☐

**Remark 8.** The result of Lemma 1 in Seuret *et al.* (2015) is generalized by Corollary 2 as $\mathbf{f}(\cdot)$ can be chosen to only have Legendre polynomials with $\varpi(\tau) = 1$. Moreover, let $\acute{\boldsymbol{f}}(\cdot)$, $\grave{\boldsymbol{f}}(\cdot)$ in (7) to contain only orthogonal functions over $[-r_1, 0]$ and $[-r_2, -r_1]$, respectively, then $\acute{\mathsf{E}}_d$ and $\grave{\mathsf{E}}_\delta$ in (12) follows the property in (41) with $\varpi(\tau) = 1$.

**Remark 9.** In (27), $\mathbf{f}(\cdot)$ can be interpreted as to approximate $\mathbf{g}(\cdot)$. By letting $\mathbf{f}(\tau) = \acute{\boldsymbol{f}}(\tau)$ and $\mathbf{g}(\tau) = \boldsymbol{\varphi}_1(\tau)$ with $\varpi(\tau) = 1$ in Lemma 3, then we have $\mathcal{E}_d = \acute{\mathsf{E}}_d$ where the matrix $\acute{\mathsf{E}}_d$ is defined in (12). Similar procedures can be applied with $\mathbf{f}(\tau) = \grave{\boldsymbol{f}}(\tau)$ and $\mathbf{g}(\tau) = \boldsymbol{\varphi}_2(\tau)$ and $\varpi(\tau) = 1$. Furthermore, if $\mathbf{f}(\cdot)$ contain only functions which are orthogonal with respect to $\varpi(\cdot)$, then the behavior of $\mathcal{E}_d$ can be quantitatively characterized with respect to $d$, which will be elaborated in the following corollary. Note that (29) holds for any $\mathsf{A}$ as long as (26) is satisfied, even if $\mathsf{A}$ is not defined as $\mathsf{A} = \int_{\mathcal{K}} \varpi(\tau)\mathbf{g}(\tau)\mathbf{f}^\top(\tau)\mathrm{d}\tau\,\mathcal{F}_d \in \mathbb{R}^{\delta\times d}$. This is an important conclusion as it infers that the error matrices $\acute{\mathsf{E}}_d$ and $\grave{\mathsf{E}}_\delta$ in (12) are invertible since (8) hold.

## 4. Main results on dissipativity and stability analysis

The main result on the dissipativity and stability analysis of (1) is presented in Theorem 1 where the condition for the dissipativity and stability analysis of (1) is denoted in terms of LMIs. Moreover, we will also show in Corollary 3, 4 that the resulting condition in Theorem 1 can exhibit a hierarchical pattern if certain perquisites are satisfied.

**Theorem 1.** *Suppose that all functions and the parameters in* (3)–(12) *are given with* $\mu_1;\mu_2 \in \mathbb{N}_0$ *and* $d;\delta \in \mathbb{N}$. *Assume also that there exist* $\acute{\boldsymbol{g}}(\cdot) \in \mathbb{C}^1(\mathbb{R} \,⨾\, \mathbb{R}^{p_1})$, $\grave{\boldsymbol{g}}(\cdot) \in \mathbb{C}^1(\mathbb{R} \,⨾\, \mathbb{R}^{p_2})$ *and* $N_1 \in \mathbb{R}^{p_1\times d}$, $N_2 \in \mathbb{R}^{p_2\times\delta}$ *such that*

$$
\begin{aligned}
&(\tau+r_1)\frac{\mathsf{d}\acute{\boldsymbol{g}}(\tau)}{\mathsf{d}\tau} = N_1\acute{\boldsymbol{f}}(\tau) && (\tau+r_2)\frac{\mathsf{d}\grave{\boldsymbol{g}}(\tau)}{\mathsf{d}\tau} = N_2\grave{\boldsymbol{f}}(\tau)\\
&\mathbb{S}^{p_1} \ni \acute{\mathsf{G}}_{p_1}^{-1} = \int_{-r_1}^{0}(\tau+r_1)\acute{\boldsymbol{g}}(\tau)\acute{\boldsymbol{g}}^\top(\tau)\mathsf{d}\tau \succ 0 && \mathbb{S}^{p_2} \ni \grave{\mathsf{G}}_{p_2}^{-1} = \int_{-r_2}^{-r_1}(\tau+r_2)\grave{\boldsymbol{g}}(\tau)\grave{\boldsymbol{g}}^\top(\tau)\mathsf{d}\tau \succ 0.
\end{aligned}
\tag{45}
$$

*Given* $r_2 > r_1 > 0$, *then the delay system* (16) *with the supply rate function* (24) *is dissipative and the origin of* (16) *is globally asymptotically stable with* $\boldsymbol{w}(t) \equiv \mathbf{0}_q$, *if there exist* $P \in \mathbb{S}^l$ *and* $Q_1;Q_2;R_1;R_2;S_1;S_2;U_1;U_2 \in \mathbb{S}^\nu$ *such that the inequalities*

$$
\begin{aligned}
\mathbf{P} := P + \left(\mathsf{O}_{n+2\nu} \oplus \left[\acute{\mathsf{F}}_d \otimes Q_1\right] \oplus \left[\grave{\mathsf{F}}_\delta \otimes Q_2\right]\right) + \Pi^\top\left(G_1^\top\acute{\Phi}_{\kappa_1}G_1 \otimes S_1 + G_2^\top\grave{\Phi}_{\kappa_2}G_2 \otimes S_2\right)\Pi\\
+ \Pi^\top\left(H_1^\top\acute{\mathsf{G}}_{p_1}H_1 \otimes U_1 + H_2^\top\grave{\mathsf{G}}_{p_2}H_2 \otimes U_2\right)\Pi \succ 0,
\end{aligned}
\tag{46}
$$

$$
Q_1 \succeq 0,\ \ Q_2 \succeq 0,\ \ R_1 \succeq 0,\ \ R_2 \succeq 0,\ \ S_1 \succeq 0,\ \ S_2 \succeq 0,\ \ U_1 \succeq 0,\ \ U_2 \succeq 0,
\tag{47}
$$

$$
\widetilde{\mathbf{\Omega}} = \begin{bmatrix} J_1 & \mathsf{O}_{m\times\nu} & \widetilde{J}\Sigma\\ * & -S_1 - r_1U_1 & (S_1 + r_1U_1)(A_6\mathbf{A} + Y)\\ * & * & \mathbf{\Omega}\end{bmatrix} \prec 0
\tag{48}
$$

*hold, where the positive definite matrices* $\acute{\mathsf{F}}_d$, $\grave{\mathsf{F}}_\delta$, $\acute{\Phi}_{\kappa_1}$ *and* $\grave{\Phi}_{\kappa_2}$ *are given in* (5) *and* (6), *and the parameters* $\mathbf{A}$ *and* $\Sigma$ *have been defined in* (17)–(19). *Moreover,*

$$
\Pi := \begin{bmatrix} \Xi & \\ \mathsf{O}_{(2\nu+\varrho\nu)\times n} & I_{2\nu+\varrho\nu}\end{bmatrix},\ \ Y := \begin{bmatrix} A_7 & A_8 & \mathsf{O}_{\nu\times(q+l+\mu\nu)}\end{bmatrix}
\tag{49}
$$

*with* $\varrho = d+\delta$ *and* $l = n + 2\nu + \varrho\nu$ *and* $\mu = \mu_1 + \mu_2$, *and*

$$
\begin{aligned}
\mathbf{\Omega} := \mathbf{Sy}\left(\Theta_2^\top P\Theta_1 - \begin{bmatrix}\mathsf{O}_{(2\nu+q+l+\mu\nu)\times 2\nu} & \Sigma^\top J_2 & \mathsf{O}_{(2\nu+q+l+\mu\nu)\times(l+\mu\nu)}\end{bmatrix}\right)\\
- \left(\mathsf{O}_{q+2\nu} \oplus \left[\Pi^\top\left(G_1^\top\acute{\Phi}_{\kappa_1}G_1 \otimes U_1 + G_2^\top\grave{\Phi}_{\kappa_2}G_2 \otimes U_2\right)\Pi\right] \oplus \mathsf{O}_{\mu\nu}\right)\\
- \Bigg([S_1 - S_2 - r_3U_2] \oplus S_2 \oplus J_3 \oplus \mathsf{O}_n \oplus [Q_1 - Q_2 - r_3R_2] \oplus Q_2 \oplus \left[\acute{\mathsf{F}}_d \otimes R_1\right] \oplus \left[\grave{\mathsf{F}}_\delta \otimes R_2\right]\\
\oplus \left[\acute{\mathsf{E}}_d \otimes R_1\right] \oplus \left[\grave{\mathsf{E}}_\delta \otimes R_2\right]\Bigg) + \begin{bmatrix}\mathsf{O}_{\nu\times(2\nu+q)} & \Xi & \mathsf{O}_{\nu\times\nu\mu}\end{bmatrix}^\top (Q_1 + r_1R_1)\begin{bmatrix}\mathsf{O}_{\nu\times(2\nu+q)} & \Xi & \mathsf{O}_{\nu\times\nu\mu}\end{bmatrix}
\end{aligned}
\tag{50}
$$

*where*

$$
G_1 = \begin{bmatrix}\acute{\boldsymbol{\phi}}(0) & -\acute{\boldsymbol{\phi}}(-r_1) & \mathbf{0}_{\kappa_1} & -M_3 & \mathsf{O}_{\kappa_1\times d}\end{bmatrix} \quad G_2 = \begin{bmatrix}\mathbf{0}_{\kappa_2} & \grave{\boldsymbol{\phi}}(-r_1) & -\grave{\boldsymbol{\phi}}(-r_2) & \mathsf{O}_{\kappa_2\times\delta} & -M_4\end{bmatrix}
\tag{51}
$$

$$
G_3 = \begin{bmatrix}\acute{\boldsymbol{f}}(0) & -\acute{\boldsymbol{f}}(-r_1) & \mathbf{0}_d & -M_1 & \mathsf{O}_d\end{bmatrix} \quad G_4 = \begin{bmatrix}\mathbf{0}_\delta & \grave{\boldsymbol{f}}(-r_1) & -\grave{\boldsymbol{f}}(-r_2) & \mathsf{O}_\delta & -M_2\end{bmatrix},
\tag{52}
$$

$$
H_1 = \begin{bmatrix} r_1\acute{\boldsymbol{g}}(0) & \mathbf{0}_{p_1} & \mathbf{0}_{p_1} & -N_1 & \mathsf{O}_{p_1}\end{bmatrix} \quad H_2 = \begin{bmatrix}\mathbf{0}_{p_2} & (r_2 - r_1)\grave{\boldsymbol{g}}(-r_1) & \mathbf{0}_{p_2} & \mathsf{O}_{p_2} & -N_2\end{bmatrix}
\tag{53}
$$

$$
\Theta_1 := \begin{bmatrix} & \mathbf{A} & \\ I_{2\nu} & \mathsf{O}_{2\nu\times(q+l+\mu\nu)} & \\ \mathsf{O}_{d\nu\times(2\nu+q)} & (G_3 \otimes I_\nu)\Pi & \mathsf{O}_{d\nu\times\nu\mu}\\ \mathsf{O}_{\delta\nu\times(2\nu+q)} & (G_4 \otimes I_\nu)\Pi & \mathsf{O}_{\delta\nu\times\nu\mu}\end{bmatrix} \quad \Theta_2 := \begin{bmatrix}\mathsf{O}_{l\times(2\nu+q)} & I_l & \mathsf{O}_{l\times\mu\nu}\end{bmatrix}.
\tag{54}
$$

*Proof.* Given $r_2 > r_1 > 0$, we consider the following Liapunov-Krasovskii functional

$$
\begin{aligned}
v(\boldsymbol{x}(t),\boldsymbol{y}(t+\cdot)) &= \boldsymbol{\eta}^\top(t)P\boldsymbol{\eta}(t) + \int_{-r_1}^{0}\boldsymbol{y}^\top(t+\tau)\left[Q_1+(\tau+r_1)R_1\right]\boldsymbol{y}(t+\tau)\mathsf{d}\tau \\
&+\int_{-r_2}^{-r_1}\boldsymbol{y}^\top(t+\tau)\left[Q_2+(\tau+r_2)R_2\right]\boldsymbol{y}(t+\tau)\mathsf{d}\tau+\int_{-r_1}^{0}\dot{\boldsymbol{y}}^\top(t+\tau)\left[S_1+(\tau+r_1)U_1\right]\dot{\boldsymbol{y}}(t+\tau)\mathsf{d}\tau \\
&+\int_{-r_2}^{-r_1}\dot{\boldsymbol{y}}^\top(t+\tau)\left[S_2+(\tau+r_2)U_2\right]\dot{\boldsymbol{y}}(t+\tau)\mathsf{d}\tau \qquad (55)
\end{aligned}
$$

to be constructed to prove the statements in Theorem 1, where $\boldsymbol{x}(t)$ and $\mathbf{y}_t(\cdot)$ here follow the same definition in (23). Moreover,

$$
\boldsymbol{\eta}(t) := \mathbf{Col}\left[\boldsymbol{x}(t),\ \boldsymbol{y}(t-r_1),\ \boldsymbol{y}(t-r_2),\ \int_{-r_1}^{0}\acute{F}_d(\tau)\boldsymbol{y}(t+\tau)\mathsf{d}\tau,\ \int_{-r_2}^{-r_1}\grave{F}_\delta(\tau)\boldsymbol{y}(t+\tau)\mathsf{d}\tau\right] \qquad (56)
$$

with $\acute{F}_d(\tau)$ and $\grave{F}_\delta(\tau)$ defined in (20), and the matrix parameters in (55) are defined as $P\in\mathbb{S}^l$ and $Q_1;Q_2;R_1;R_2;S_1;S_2;U_1;U_2\in\mathbb{S}^\nu$ with $l:=n+2\nu+\varrho\nu$ and $\varrho:=d+\delta$. Note that since the eigenvalues of all the matrix terms $Q_1+(\tau+r_1)R_1$, $Q_2+(\tau+r_2)R_2$ $S_1+(\tau+r_1)U_1$ and $S_2+(\tau+r_2)U_2$ in (55) are bounded, thus all the quadratic integrals associated with these terms are well defined since $\mathbf{y}_t(\cdot)\in\mathcal{A}\left([-r_2,0)\,⨾\,\mathbb{R}^\nu\right)$. On the other hand, since $\mathbf{y}_t(\cdot)$, $\acute{\boldsymbol{f}}(\tau)$ and $\grave{\boldsymbol{f}}(\tau)$ are bounded, thus the integrals in (56) are well defined as well.

Firstly, we prove that the existence of the feasible solutions of (47) and (48) infers that (55) satisfies both (22) and (23). Subsequently, we show that the existence of the feasible solutions of (46) and (47) infers that (55) satisfies (21). The existence of the upper bound of $v(\boldsymbol{x}(t),\mathbf{y}_t(\cdot))$ can be independently proved without considering the inequalities (46)–(48).

Let $t_0\in\mathbb{R}$, differentiate $v(\boldsymbol{x}(t),\mathbf{y}_t(\cdot))$ along the trajectory of (16) and consider (24), it produces

$$
\begin{aligned}
\forall t\geq t_0,\quad \dot{v}(\boldsymbol{x}(t),\mathbf{y}_t(\cdot)) - s(\boldsymbol{z}(t),\boldsymbol{w}(t)) &= \boldsymbol{\vartheta}^\top(t)\,\mathbf{Sy}\left(\Theta_2^\top P\Theta_1\right)\boldsymbol{\vartheta}(t)+\boldsymbol{y}^\top(t)\left(Q_1+r_1R_1\right)\boldsymbol{y}(t) \\
&+\boldsymbol{y}^\top(t-r_1)\left(Q_2+r_3R_2-Q_1\right)\boldsymbol{y}(t-r_1)-\boldsymbol{y}^\top(t-r_2)Q_2\boldsymbol{y}(t-r_2)+\dot{\boldsymbol{y}}^\top(t)\left(S_1+r_1U_1\right)\dot{\boldsymbol{y}}(t) \\
&+\dot{\boldsymbol{y}}^\top(t-r_1)\left(S_2+r_3U_2-S_1\right)\dot{\boldsymbol{y}}(t-r_1)-\dot{\boldsymbol{y}}^\top(t-r_2)S_2\dot{\boldsymbol{y}}(t-r_2)-\int_{-r_1}^{0}\boldsymbol{y}^\top(t+\tau)R_1\boldsymbol{y}(t+\tau)\mathsf{d}\tau \\
&-\int_{-r_2}^{-r_1}\boldsymbol{y}^\top(t+\tau)R_2\boldsymbol{y}(t+\tau)\mathsf{d}\tau-\int_{-r_1}^{0}\dot{\boldsymbol{y}}^\top(t+\tau)U_1\dot{\boldsymbol{y}}(t+\tau)\mathsf{d}\tau-\int_{-r_2}^{-r_1}\dot{\boldsymbol{y}}^\top(t+\tau)U_2\dot{\boldsymbol{y}}(t+\tau)\mathsf{d}\tau \\
&-\boldsymbol{w}^\top(t)J_3\boldsymbol{w}(t)-\boldsymbol{\vartheta}^\top(t)\left[\Sigma^\top\widetilde{J}^\top J_1^{-1}\widetilde{J}\Sigma+\mathbf{Sy}\left(\left[\mathsf{O}_{(2\nu+q+l+\mu\nu)\times 2\nu}\quad \Sigma^\top J_2\quad \mathsf{O}_{(2\nu+q+l+\mu\nu)\times(l+\mu\nu)}\right]\right)\right]\boldsymbol{\vartheta}(t) \qquad (57)
\end{aligned}
$$

where $\boldsymbol{\vartheta}(t)$ and $\Theta_1;\Theta_2$ have been defined in (20) and (54), respectively, and the matrices $G_3$ and $G_4$ in (52) are obtained via the relations

$$
\begin{aligned}
\int_{-r_1}^{0}\acute{F}_d(\tau)\dot{\boldsymbol{y}}(t+\tau)\mathsf{d}\tau &= \acute{F}_d(0)\boldsymbol{y}(t)-\acute{F}_d(-r_1)\boldsymbol{y}(t-r_1)-(M_1\otimes I_\nu)\int_{-r_1}^{0}\acute{F}_d(\tau)\boldsymbol{y}(t+\tau)\mathsf{d}\tau = \\
&= \left[\mathsf{O}_{d\nu\times(q+2\nu)}\quad (G_3\otimes I_\nu)\,\Pi\quad \mathsf{O}_{d\nu\times\nu\mu}\right]\boldsymbol{\vartheta}(t)
\end{aligned} \qquad (58)
$$

$$
\begin{aligned}
\int_{-r_2}^{-r_1}\grave{F}_\delta(\tau)\dot{\boldsymbol{y}}(t+\tau)\mathsf{d}\tau &= \grave{F}_\delta(-r_1)\boldsymbol{y}(t-r_1)-\grave{F}_\delta(-r_2)\boldsymbol{y}(t-r_2)-(M_2\otimes I_\nu)\int_{-r_2}^{-r_1}\grave{F}_\delta(\tau)\boldsymbol{y}(t+\tau)\mathsf{d}\tau \\
&= \left[\mathsf{O}_{\delta\nu\times(q+2\nu)}\quad (G_4\otimes I_\nu)\,\Pi\quad \mathsf{O}_{\delta\nu\times\nu\mu}\right]\boldsymbol{\vartheta}(t)
\end{aligned} \qquad (59)
$$

which themselves can be derived by using (15), (14) with (3).

To obtain a upper bound for (57), let $R_1 \succeq 0$, $R_2 \succeq 0$ so that the inequalities

$$\begin{aligned}\int_{-r_1}^{0} \boldsymbol{y}^\top(t+\tau)R_1\boldsymbol{y}(t+\tau)\mathsf{d}\tau &\geq \int_{-r_1}^{0} \boldsymbol{y}^\top(t+\tau)\acute{F}_d^\top(\tau)\mathsf{d}\tau \left(\acute{\mathsf{F}}_d \otimes R_1\right) \int_{-r_1}^{0} \acute{F}_d(\tau)\boldsymbol{y}(t+\tau)\mathsf{d}\tau \\ &+ \int_{-r_1}^{0} \boldsymbol{y}^\top(t+\tau)\acute{E}_d^\top(\tau)\mathsf{d}\tau \left(\acute{\mathsf{E}}_d \otimes R_1\right) \int_{-r_1}^{0} \acute{E}_d(\tau)\boldsymbol{y}(t+\tau)\mathsf{d}\tau \\ \int_{-r_2}^{-r_1} \boldsymbol{y}^\top(t+\tau)R_2\boldsymbol{y}(t+\tau)\mathsf{d}\tau &\geq \int_{-r_2}^{-r_1} \boldsymbol{y}^\top(t+\tau)\grave{F}_\delta^\top(\tau)\mathsf{d}\tau \left(\grave{\mathsf{F}}_\delta \otimes R_2\right) \int_{-r_2}^{-r_1} \grave{F}_\delta(\tau)\boldsymbol{y}(t+\tau)\mathsf{d}\tau \\ &+ \int_{-r_2}^{-r_1} \boldsymbol{y}^\top(t+\tau)\grave{E}_\delta^\top(\tau)\mathsf{d}\tau \left(\grave{\mathsf{E}}_\delta \otimes R_2\right) \int_{-r_2}^{-r_1} \grave{E}_\delta(\tau)\boldsymbol{y}(t+\tau)\mathsf{d}\tau\end{aligned} \tag{60}$$

can be derived from (27) with $\mathsf{f}(\tau) = \acute{\boldsymbol{f}}(\tau)$; $\mathsf{g}(\tau) = \boldsymbol{\varphi}_1(\tau)$ and $\mathsf{f}(\tau) = \grave{\boldsymbol{f}}(\tau)$; $\mathsf{g}(\tau) = \boldsymbol{\varphi}_2(\tau)$, respectively, which matches $\acute{F}_d(\tau)$; $\grave{F}_\delta(\tau)$ in (20) and the expressions in (11). Furthermore, let $U_1 \succeq 0$ and $U_2 \succeq 0$ and apply (38) to the integral terms $\int_{-r_1}^{0} \dot{\boldsymbol{y}}^\top(t+\tau)U_1\dot{\boldsymbol{y}}(t+\tau)\mathsf{d}\tau$ and $\int_{-r_2}^{-r_1} \dot{\boldsymbol{y}}^\top(t+\tau)U_2\dot{\boldsymbol{y}}(t+\tau)\mathsf{d}\tau$ with $\mathsf{f}(\tau) = \acute{\boldsymbol{\phi}}(\tau)$ and $\mathsf{f}(\tau) = \grave{\boldsymbol{\phi}}(\tau)$ in (6), respectively, and consider the expression $\boldsymbol{y}(t) = \begin{bmatrix}\mathsf{O}_{\nu\times(2\nu+q)} & \Xi & \mathsf{O}_{\nu\times\nu\mu}\end{bmatrix}\boldsymbol{\vartheta}(t)$ in (18) with (14) and (15). It produces

$$\begin{aligned}\int_{-r_1}^{0} \dot{\boldsymbol{y}}^\top(t+\tau)U_1\dot{\boldsymbol{y}}(t+\tau)\mathsf{d}\tau &\geq \int_{-r_1}^{0} \dot{\boldsymbol{y}}^\top(t+\tau)\left(\acute{\boldsymbol{\phi}}^\top(\tau)\otimes I_\nu\right)\mathsf{d}\tau \left(\acute{\Phi}_{\kappa_1}\otimes U_1\right)\int_{-r_1}^{0}\left(\acute{\boldsymbol{\phi}}(\tau)\otimes I_\nu\right)\dot{\boldsymbol{y}}(t+\tau)\mathsf{d}\tau \\ &= \boldsymbol{\vartheta}^\top(\tau)\left[\mathsf{O}_{2\nu+q}\oplus\Pi^\top\left(G_1^\top\acute{\Phi}_{\kappa_1}G_1\otimes U_1\right)\Pi\oplus\mathsf{O}_{\nu\mu}\right]\boldsymbol{\vartheta}(\tau),\end{aligned} \tag{61}$$

$$\begin{aligned}\int_{-r_2}^{-r_1} \dot{\boldsymbol{y}}^\top(t+\tau)U_2\dot{\boldsymbol{y}}(t+\tau)\mathsf{d}\tau &\geq \int_{-r_2}^{-r_1} \dot{\boldsymbol{y}}^\top(t+\tau)\left(\grave{\boldsymbol{\phi}}^\top(\tau)\otimes I_\nu\right)\mathsf{d}\tau \left(\grave{\Phi}_{\kappa_2}\otimes U_2\right)\int_{-r_2}^{-r_1}\left(\grave{\boldsymbol{\phi}}(\tau)\otimes I_\nu\right)\dot{\boldsymbol{y}}(t+\tau)\mathsf{d}\tau \\ &= \boldsymbol{\vartheta}^\top(\tau)\left[\mathsf{O}_{2\nu+q}\oplus\Pi^\top\left(G_2^\top\grave{\Phi}_{\kappa_2}G_2\otimes U_2\right)\Pi\oplus\mathsf{O}_{\nu\mu}\right]\boldsymbol{\vartheta}(\tau)\end{aligned} \tag{62}$$

where $G_1$ and $G_2$ are given in (51) which are derived by the relations

$$\begin{aligned}\int_{-r_1}^{0}\left(\acute{\boldsymbol{\phi}}(\tau)\otimes I_\nu\right)\dot{\boldsymbol{y}}(t+\tau)\mathsf{d}\tau &= \left(\acute{\boldsymbol{\phi}}(0)\otimes I_\nu\right)\boldsymbol{y}(t) - \left(\acute{\boldsymbol{\phi}}(-r_1)\otimes I_\nu\right)\boldsymbol{y}(t-r_1) \\ &- (M_3\otimes I_\nu)\int_{-r_1}^{0}\left(\acute{\boldsymbol{\phi}}(\tau)\otimes I_\nu\right)\boldsymbol{y}(t+\tau)\mathsf{d}\tau = (G_1\otimes I_\nu)\,\Pi\boldsymbol{\eta}(t) \\ &= \begin{bmatrix}\mathsf{O}_{\kappa_1\nu\times(q+2\nu)} & (G_1\otimes I_\nu)\,\Pi & \mathsf{O}_{\kappa_1\nu\times\nu\mu}\end{bmatrix}\boldsymbol{\vartheta}(t)\end{aligned} \tag{63}$$

$$\begin{aligned}\int_{-r_2}^{-r_1}\left(\grave{\boldsymbol{\phi}}(\tau)\otimes I_\nu\right)\dot{\boldsymbol{y}}(t+\tau)\mathsf{d}\tau &= \left(\grave{\boldsymbol{\phi}}(-r_1)\otimes I_\nu\right)\boldsymbol{y}(t-r_1) - \left(\grave{\boldsymbol{\phi}}(-r_2)\otimes I_\nu\right)\boldsymbol{y}(t-r_2) \\ &- (M_4\otimes I_\nu)\int_{-r_2}^{-r_1}\left(\grave{\boldsymbol{\phi}}(\tau)\otimes I_\nu\right)\boldsymbol{y}(t+\tau)\mathsf{d}\tau = (G_2\otimes I_\nu)\,\Pi\boldsymbol{\eta}(t) \\ &= \begin{bmatrix}\mathsf{O}_{\kappa_2\nu\times(q+2\nu)} & (G_2\otimes I_\nu)\,\Pi & \mathsf{O}_{\kappa_2\nu\times\nu\mu}\end{bmatrix}\boldsymbol{\vartheta}(t).\end{aligned} \tag{64}$$

Now applying (60)–(62) with (47) to (57) yields

$$\begin{aligned}&\forall t\geq t_0,\;\; \dot{v}(\boldsymbol{x}(t),\boldsymbol{y}_t(\cdot)) - s(\boldsymbol{z}(t),\boldsymbol{w}(t)) \leq \\ &\boldsymbol{\vartheta}^\top(t)\left[\boldsymbol{\Omega} + (A_6\mathbf{A}+Y)^\top(S_1+r_1U_1)(A_6\mathbf{A}+Y) - \Sigma^\top\widetilde{J}^\top J_1^{-1}\widetilde{J}\Sigma\right]\boldsymbol{\vartheta}(t)\end{aligned} \tag{65}$$

where $\boldsymbol{\Omega}$ has been defined in (50). It is obvious that if $\boldsymbol{\Omega} + (A_6\mathbf{A}+Y)^\top(S_1+r_1U_1)(A_6\mathbf{A}+Y) - \Sigma^\top\widetilde{J}^\top J_1^{-1}\widetilde{J}\Sigma \prec 0$ holds with (47), we have

$$\exists\epsilon_3>0,\,\forall t\geq t_0,\;\; \dot{v}(\boldsymbol{x}(t),\boldsymbol{y}_t(\cdot)) - s(\boldsymbol{z}(t),\boldsymbol{w}(t)) \leq -\epsilon_3\|\boldsymbol{x}(t)\|_2. \tag{66}$$

Moreover, let $\boldsymbol{w}(t) \equiv \mathbf{0}_q$ and consider the structure of the quadratic term in (65) together with the properties of negative definite matrices. One can conclude that if $\boldsymbol{\Omega} + (A_6\mathbf{A} + Y)^\top (S_1 + r_1U_1)(A_6\mathbf{A} + Y) - \Sigma^\top \widetilde{J}^\top J_1^{-1} \widetilde{J}\Sigma \prec 0$ and (47) are satisfied, it infers that

$$\exists\epsilon_3 > 0, \quad \frac{\mathsf{d}^+}{\mathsf{d}t} v(\boldsymbol{x}(t), \mathbf{y}_t(\cdot))\bigg|_{t=t_0, \boldsymbol{x}(t_0)=\boldsymbol{\xi}, \mathbf{y}_{t_0}(\cdot)=\boldsymbol{\phi}(\cdot)} = \dot{v}(\boldsymbol{\xi}, \boldsymbol{\phi}(\cdot)) \le -\epsilon_3 \|\boldsymbol{\xi}\|_2 \tag{67}$$

where $\boldsymbol{x}(t)$ and $\mathbf{y}_t(\cdot)$ here follow the same definition in Lemma 2. As a result, it is obvious that (47) with $\boldsymbol{\Omega} + (A_6\mathbf{A} + Y)^\top (S_1 + r_1U_1)(A_6\mathbf{A} + Y) - \Sigma^\top \widetilde{J}^\top J_1^{-1} \widetilde{J}\Sigma \prec 0$ infers (22) and (23). Finally, applying the Schur complement to $\boldsymbol{\Omega} + (A_6\mathbf{A} + Y)^\top (S_1 + r_1U_1)(A_6\mathbf{A} + Y) - \Sigma^\top \widetilde{J}^\top J_1^{-1} \widetilde{J}\Sigma \prec 0$ with (47) and $J_1^{-1} \prec 0$ gives (48). Hence we have proved that the feasible solutions of (47) and (48) infers that (55) satisfies (22) and (23).

Now we start to prove that if (46) and (47) hold then (55) satisfies (21). Let $\|\boldsymbol{\psi}(\cdot)\|_\infty := \sup_{-r_2 \le r \le 0} \|\boldsymbol{\psi}(\tau)\|_2$ and $\|\boldsymbol{\psi}(\cdot)\|_2^2 := \int_{-r_2}^0 \boldsymbol{\psi}^\top(\tau)\boldsymbol{\psi}(\tau)\mathsf{d}\tau$. Given the structure of (55) with $t = t_0$, it follows that $\exists\lambda; \eta > 0:$

$$\begin{aligned}
v(\boldsymbol{x}(t_0), \mathbf{y}_{t_0}(\cdot)) &= v(\boldsymbol{\xi}, \boldsymbol{\psi}(\cdot)) \le \boldsymbol{\eta}^\top(t_0)\lambda\boldsymbol{\eta}(t_0) + \lambda \int_{-r_2}^0 \begin{bmatrix} \boldsymbol{\psi}^\top(\tau) & \dot{\boldsymbol{\psi}}^\top(\tau) \end{bmatrix} \begin{bmatrix} \boldsymbol{\psi}(\tau) \\ \dot{\boldsymbol{\psi}}(\tau) \end{bmatrix} \mathsf{d}\tau \le \lambda\|\boldsymbol{\xi}\|_2^2 \\
&+ (2\lambda + \lambda r_2)\|\boldsymbol{\psi}(\cdot)\|_\infty^2 + \lambda r_2 \|\dot{\boldsymbol{\psi}}(\cdot)\|_2^2 + \lambda \int_{-r_1}^0 \boldsymbol{\psi}^\top(\tau)\acute{F}_d^\top(\tau)\mathsf{d}\tau \int_{-r_1}^0 \acute{F}_d(\tau)\boldsymbol{\psi}(\tau)\mathsf{d}\tau \\
&+ \lambda \int_{-r_2}^{-r_1} \boldsymbol{\psi}^\top(\tau)\grave{F}_\delta^\top(\tau)\mathsf{d}\tau \int_{-r_2}^{-r_1} \grave{F}_\delta(\tau)\boldsymbol{\psi}(\tau)\mathsf{d}\tau \le \lambda\|\boldsymbol{\xi}\|_2^2 + (2\lambda + \lambda r_2)\|\boldsymbol{\psi}(\cdot)\|_\infty^2 \\
&+ \lambda r_2 \|\dot{\boldsymbol{\psi}}(\cdot)\|_2^2 + \int_{-r_1}^0 \boldsymbol{\psi}^\top(\tau)\acute{F}_d^\top(\tau)\mathsf{d}\tau \left(\eta\acute{\mathsf{F}}_d \otimes I_\nu\right) \int_{-r_1}^0 \acute{F}_d(\tau)\boldsymbol{\psi}(\tau)\mathsf{d}\tau \\
&+ \int_{-r_2}^{-r_1} \boldsymbol{\psi}^\top(\tau)\grave{F}_\delta^\top(\tau)\mathsf{d}\tau \left(\eta\grave{\mathsf{F}}_\delta \otimes I_\nu\right) \int_{-r_2}^{-r_1} \grave{F}_\delta(\tau)\boldsymbol{\psi}(\tau)\mathsf{d}\tau \le \lambda\|\boldsymbol{\xi}\|_2^2 + (2\lambda + \lambda r_2)\|\boldsymbol{\psi}(\cdot)\|_\infty^2 \\
&+ \lambda r_2 \|\dot{\boldsymbol{\psi}}(\cdot)\|_2^2 + \eta \int_{-r_1}^0 \boldsymbol{\psi}^\top(\tau)\boldsymbol{\psi}(\tau)\mathsf{d}\tau + \eta \int_{-r_2}^{-r_1} \boldsymbol{\psi}^\top(\tau)\boldsymbol{\psi}(\tau)\mathsf{d}\tau = \lambda\|\boldsymbol{\xi}\|_2^2 \\
&+ (2\lambda + \lambda r_2 + \eta r_2)\|\boldsymbol{\psi}(\cdot)\|_\infty^2 + \lambda r_2\|\dot{\boldsymbol{\psi}}(\cdot)\|_2^2 \le (2\lambda + \lambda r_2 + \eta r_2)\left(\|\boldsymbol{\xi}\|_2^2 + \|\boldsymbol{\psi}(\cdot)\|_\infty^2 + \|\dot{\boldsymbol{\psi}}(\cdot)\|_2^2\right) \\
&\le (2\lambda + \lambda r_2 + \eta r_2)\left[\|\boldsymbol{\xi}\|_2^2 + \left(\|\boldsymbol{\psi}(\cdot)\|_\infty + \|\dot{\boldsymbol{\psi}}(\cdot)\|_2\right)^2\right] \\
&\le (4\lambda + 2\lambda r_2 + 2\eta r_2)\left[\|\boldsymbol{\xi}\|_2 \vee \left(\|\boldsymbol{\psi}(\cdot)\|_\infty + \|\dot{\boldsymbol{\psi}}(\cdot)\|_2\right)\right]^2
\end{aligned} \tag{68}$$

for any initial condition $\boldsymbol{\xi} \in \mathbb{R}^n$ and $\boldsymbol{\psi}(\cdot) \in \mathcal{A}([-r_2, 0) ⨾ \mathbb{R}^\nu)$ in (1), which is derived via (38) and the property of quadratic forms: $\forall X \in \mathbb{S}^n, \exists\lambda > 0 : \forall \mathbf{x} \in \mathbb{R}^n \setminus \{\mathbf{0}\}, \mathbf{x}^\top(\lambda I_n - X)\mathbf{x} > 0$. Then (68) shows that (55) satisfies the rightmost inequality in (21).

Now assume the inequalities in (47) are satisfied. Apply (38) with appropriate $\mathbf{f}(\cdot)$ with $\varpi(\tau) = 1$ to the integrals in (55) at $t = t_0$ and consider the initial conditions in (1), we have

$$\begin{aligned}
\int_{-r_1}^0 \boldsymbol{\psi}^\top(\tau)Q_1\boldsymbol{\psi}(\tau)\mathsf{d}\tau &\ge \int_{-r_1}^0 \boldsymbol{\psi}^\top(\tau)\acute{F}_d^\top(\tau)\mathsf{d}\tau \left(\acute{\mathsf{F}}_d \otimes Q_1\right) \int_{-r_1}^0 \acute{F}_d(\tau)\boldsymbol{\psi}(\tau)\mathsf{d}\tau, \\
\int_{-r_2}^{-r_1} \boldsymbol{\psi}^\top(\tau)Q_2\boldsymbol{\psi}(\tau)\mathsf{d}\tau &\ge \int_{-r_2}^{-r_1} \boldsymbol{\psi}^\top(\tau)\grave{F}_\delta^\top(\tau)\mathsf{d}\tau \left(\grave{\mathsf{F}}_\delta \otimes Q_2\right) \int_{-r_2}^{-r_1} \grave{F}_\delta(\tau)\boldsymbol{\psi}(\tau)\mathsf{d}\tau
\end{aligned} \tag{69}$$

and

$$\int_{-r_1}^0 \dot{\boldsymbol{\psi}}^\top(\tau)S_1\dot{\boldsymbol{\psi}}(\tau)\mathsf{d}\tau \ge \int_{-r_1}^0 \dot{\boldsymbol{\psi}}^\top(\tau)\left(\acute{\boldsymbol{\phi}}^\top(\tau) \otimes I_\nu\right)\mathsf{d}\tau \left(\acute{\mathsf{\Phi}}_{\kappa_1} \otimes S_1\right) \int_{-r_1}^0 \left(\acute{\boldsymbol{\phi}}(\tau) \otimes I_\nu\right)\dot{\boldsymbol{\psi}}(\tau)\mathsf{d}\tau$$

$$= \boldsymbol{\eta}^\top(t_0)\Pi^\top \left(G_1^\top \acute{\Phi}_{\kappa_1} G_1 \otimes S_1\right) \Pi\boldsymbol{\eta}(t_0), \quad (70)$$

$$\int_{-r_2}^{-r_1} \dot{\boldsymbol{\psi}}^\top(\tau) S_2 \dot{\boldsymbol{\psi}}(\tau)\mathsf{d}\tau \geq \int_{-r_2}^{-r_1} \dot{\boldsymbol{\psi}}^\top(\tau) \left(\grave{\boldsymbol{\phi}}^\top(\tau) \otimes I_\nu\right) \mathsf{d}\tau \left(\grave{\Phi}_{\kappa_2} \otimes S_2\right) \int_{-r_2}^{-r_1} \left(\grave{\boldsymbol{\phi}}(\tau) \otimes I_\nu\right) \dot{\boldsymbol{\psi}}(\tau)\mathsf{d}\tau$$
$$= \boldsymbol{\eta}^\top(t_0)\Pi^\top \left(G_2^\top \grave{\Phi}_{\kappa_2} G_2 \otimes S_2\right) \Pi\boldsymbol{\eta}(t_0) \quad (71)$$

which are derived via the relations in (63) and (64). Furthermore, apply (27) again with appropriate weight functions to the integrals $\int_{-r_1}^{0}(r_1+\tau)\dot{\boldsymbol{y}}^\top(t+\tau)U_1\dot{\boldsymbol{y}}(t+\tau)\mathsf{d}\tau$ and $\int_{-r_2}^{-r_1}(r_2+\tau)\dot{\boldsymbol{y}}^\top(t+\tau)U_2\dot{\boldsymbol{y}}(t+\tau)\mathsf{d}\tau$ for $t = t_0$ in (55) with $\mathbf{f}(\tau) = \acute{\boldsymbol{g}}(\tau)$, $\mathbf{f}(\tau) = \grave{\boldsymbol{g}}(\tau)$, respectively. Then it yields

$$\begin{aligned}
&\int_{-r_1}^{0} (r_1+\tau)\dot{\boldsymbol{\psi}}^\top(\tau) U_1 \dot{\boldsymbol{\psi}}(\tau)\mathsf{d}\tau \geq [*] \left(\acute{\mathsf{G}}_{p_1} \otimes U_1\right) \int_{-r_1}^{0} (\tau + r_1) \left(\acute{\boldsymbol{g}}(\tau) \otimes I_\nu\right) \dot{\boldsymbol{\psi}}(\tau)\mathsf{d}\tau \\
&\qquad = \boldsymbol{\eta}^\top(t_0)\Pi^\top \left(H_1^\top \acute{\mathsf{G}}_{p_1} H_1 \otimes U_1\right) \Pi\boldsymbol{\eta}(t_0) \\
&\int_{-r_2}^{-r_1} (r_2+\tau)\dot{\boldsymbol{\psi}}^\top(\tau) U_2 \dot{\boldsymbol{\psi}}(\tau)\mathsf{d}\tau \geq [*] \left(\grave{\mathsf{G}}_{p_2} \otimes U_2\right) \int_{-r_2}^{-r_1} (\tau + r_2) \left(\grave{\boldsymbol{g}}(\tau) \otimes I_\nu\right) \dot{\boldsymbol{\psi}}(\tau)\mathsf{d}\tau \\
&\qquad = \boldsymbol{\eta}^\top(t_0)\Pi^\top \left(H_2^\top \grave{\mathsf{G}}_{p_2} H_2 \otimes U_2\right) \Pi\boldsymbol{\eta}(t_0)
\end{aligned} \quad (72)$$

for any initial condition $\boldsymbol{\xi} \in \mathbb{R}^n$ and $\boldsymbol{\psi}(\cdot) \in \mathcal{A}\left([-r_2, 0) ⨾ \mathbb{R}^\nu\right)$ in (1), where $H_1$ and $H_2$ are given in (53) and obtained by the relations

$$\int_{-r_1}^{0} (\tau + r_1) \left(\acute{\boldsymbol{g}}(\tau) \otimes I_\nu\right) \dot{\boldsymbol{\psi}}(\tau)\mathsf{d}\tau = r_1 \left(\acute{\boldsymbol{g}}(0) \otimes I_\nu\right) \boldsymbol{\psi}(0) - (N_1 \otimes I_\nu) \int_{-r_1}^{0} \left(\acute{\boldsymbol{f}}(\tau) \otimes I_\nu\right) \boldsymbol{\psi}(\tau)\mathsf{d}\tau$$
$$= (H_1 \otimes I_\nu)\, \boldsymbol{\eta}(t_0) \quad (73)$$

$$\begin{aligned}
&\int_{-r_2}^{-r_1} (\tau + r_2) \left(\grave{\boldsymbol{g}}(\tau) \otimes I_\nu\right) \dot{\boldsymbol{\psi}}(\tau)\mathsf{d}\tau = (r_2 - r_1) \left(\grave{\boldsymbol{g}}(-r_1) \otimes I_\nu\right) \boldsymbol{\psi}(-r_1) \\
&\qquad - (N_2 \otimes I_\nu) \int_{-r_2}^{-r_1} \left(\grave{\boldsymbol{f}}(\tau) \otimes I_\nu\right) \boldsymbol{\psi}(\tau)\mathsf{d}\tau = (H_2 \otimes I_\nu)\, \boldsymbol{\eta}(t_0)
\end{aligned} \quad (74)$$

via (45) and the properties of Kronecker product in (14) and (15).

With (47), utilizing (69)–(72) to (55) with $t = t_0$ and considering the initial conditions in (1) can conclude that (21) is satisfied if (46) and (47) hold. This shows that feasible solutions of (46)–(48) infers the existence of the functional in (55) satisfying (21)–(23). This finishes the proof. □

**Remark 10.** By allowing $m, q$ to be zero, Theorem 1 can cope with the problem of conducting stability analysis without performance requirements. Moreover, if $\acute{\boldsymbol{f}}(\cdot)$ and $\grave{\boldsymbol{f}}(\cdot)$ contain only Legendre polynomials, then Theorem 1 with (13) generalizes the two delay channel version of the stability results in Seuret *et al.* (2015). (Note that the method in Seuret *et al.* (2015) only deals with systems with a single delay channel)

**Remark 11.** If one wants to increase the values of $d$ and $\delta$ in (55) to incorporate more functions in the distributed delay terms in (56), then extra zeros need to be introduced to the coefficient matrices $A_4, A_5$ and $C_4, C_5$ in (7) in order to make (55) consistent with (7). In conclusion, there are no upper bound on the values of $d$ and $\delta$. Finally, (55) generalizes the Krasovskii functional in Seuret *et al.* (2015) which only consider Legendre polynomials for the integral terms in (56).

**Remark 12.** If the condition in (45) is not imposed on $\acute{\boldsymbol{f}}(\cdot)$ and $\grave{\boldsymbol{f}}(\cdot)$ then dissipative conditions can still be derived but the inequalities in (72) can no longer be considered. In that case, the constraints (47) and (48) remain the same, and (46) is changed into

$$P + \left(\mathsf{O}_{n+2\nu} \oplus \left[\acute{\mathsf{F}}_d \otimes Q_1\right] \oplus \left[\grave{\mathsf{F}}_\delta \otimes Q_2\right]\right) + \Pi^\top \left(G_1^\top \acute{\Phi}_{\kappa_1} G_1 \otimes S_1 + G_2^\top \grave{\Phi}_{\kappa_2} G_2 \otimes S_2\right) \Pi \succ 0. \quad (75)$$

**Remark 13.** Note that the position of the error matrices $\acute{\mathsf{E}}_d$ and $\grave{\mathsf{E}}_\delta$ in $\widetilde{\boldsymbol{\Omega}} \prec 0$ in (48) may cause numerical problem if the eigenvalues of $\acute{\mathsf{E}}_d$ and $\grave{\mathsf{E}}_\delta$ are too small. To circumvent this potential issue, we can apply congruence transformations to $\widetilde{\boldsymbol{\Omega}} \prec 0$ which concludes that $\widetilde{\boldsymbol{\Omega}} \prec 0$ holds if and only if

$$[*]\widetilde{\boldsymbol{\Omega}}\left[I_{m+q+n+5\nu+\varrho\nu} \oplus \left(\eta_1 \acute{\mathsf{E}}_d^{-\frac{1}{2}} \otimes I_\nu\right) \oplus \left(\eta_2 \grave{\mathsf{E}}_\delta^{-\frac{1}{2}} \otimes I_\nu\right)\right] \prec 0 \tag{76}$$

holds where $\eta_1; \eta_2 \in \mathbb{R}$ are given values. Note that the diagonal elements of the transformed matrix in (76) are no longer associated with the error terms appear at off-diagonal elements, hence one can use the inequality (76) instead of (48).

**Remark 14.** The assumption of $r_2 > r_1 > 0$ in Theorem 1 indicates that there are no obvious redundant matrix parameters in (55) since two genuine delay channels are considered therein and (56) and (20) contain no zeros vectors. With $r_1 = 0$ or $r_2 = r_1$, one only need to consider one delay channel thus the corresponding (56), (20) and (55) can be simplified. Note that we do not present the corresponding dissipativity and stability condition for $r_1 = 0$ or $r_2 = r_1$ in this paper since it can be easily derived based on the proof of Theorem 1 with a simplified (55).

In the following corollary, we show that a hierarchy of the stability condition in Theorem 1 can be established with respect to $\check{\boldsymbol{\phi}}(\cdot)$ and its dimension under certain conditions.

**Corollary 3.** *Let all the functions and the parameters in* (3)*–*(12) *be given where* $\grave{\boldsymbol{\phi}}(\tau) := \mathbf{Col}_{i=1}^{\kappa_2} \grave{\phi}_i(\tau)$ *with* $\{\grave{\phi}_i(\cdot)\}_{i=1}^{\kappa_2} \subset \{\grave{\phi}_i(\cdot)\}_{i=1}^{\infty} \subset \mathbb{C}^1([-r_2, -r_1] \,⨾\, \mathbb{R})$ *satisfying*

$$\exists \varkappa \in \mathbb{N},\ \forall \kappa_2 \in \{j \in \mathbb{N} : j \leq \varkappa\},\ \exists! M_4 \in \mathbb{R}^{\kappa_2 \times \delta},\ \frac{\mathsf{d}}{\mathsf{d}\tau} \mathbf{Col}_{i=1}^{\kappa_2} \grave{\phi}_i(\tau) = M_4 \grave{\boldsymbol{f}}(\tau) \tag{77}$$

$$\forall \kappa_2 \in \mathbb{N},\ \grave{\Phi}_{\kappa_2} = \int_{-r_2}^{-r_1} \mathbf{Col}_{i=1}^{\kappa_2} \grave{\phi}_i(\tau) \mathbf{Row}_{i=1}^{\kappa_2} \grave{\phi}_i(\tau) \mathsf{d}\tau = \bigoplus_{j=1}^{\kappa_2} \grave{\varphi}_j,\quad \grave{\varphi}_j^{-1} = \int_{-r_2}^{-r_1} \grave{\phi}_j^2(\tau) \mathsf{d}\tau. \tag{78}$$

*Now given* $\acute{\boldsymbol{g}}(\cdot)$, $\grave{\boldsymbol{g}}(\cdot)$ *and* $N_1$, $N_2$ *in Theorem 1, we have*

$$\forall \kappa_2 \in \{j \in \mathbb{N} : j \leq \varkappa\},\ \mathcal{G}_{\kappa_2} \subseteq \mathcal{G}_{\kappa_2+1} \tag{79}$$

*where* $\varkappa \in \mathbb{N}$ *is given and*

$$\mathcal{G}_{\kappa_2} := \left\{(r_1, r_2) \,\middle|\, r_1 > 0, r_2 > r_1 \ \ \& \ \ (46)\text{–}(48) \text{ hold } \ \& \ \ P \in \mathbb{S}^l, Q_1; Q_2; R_1; R_2; S_1; S_2; U_1; U_2 \in \mathbb{S}^\nu\right\}$$

*with* $l := n + 2\nu + (d + \delta)\nu$*.*

*Proof.* Given $r_2 > r_1 > 0$ and all the parameters in (3)–(7) and (11), (12), let $\mathbf{Col}(r_1, r_2) \in \mathcal{G}_{\kappa_2}$ with $\mathcal{G}_{\kappa_2} \neq \varnothing$ which infers that there exist feasible solutions for (46)–(48). Consider the situation when the dimensions and elements of $\acute{\boldsymbol{f}}(\tau)$, $\grave{\boldsymbol{f}}(\tau)$, $\acute{\boldsymbol{\phi}}(\tau)$, $\acute{\boldsymbol{g}}(\tau)$ and $\grave{\boldsymbol{g}}(\tau)$ are all fixed, and let $P \in \mathbb{S}^l$ and $Q_1; Q_2; R_1; R_2; S_1; S_2; U_1; U_2 \in \mathbb{S}^\nu$ to be a given feasible solution for $\mathbf{P}_{\kappa_2} \succ 0$, (47) and $\widetilde{\boldsymbol{\Omega}}_{\kappa_2} \prec 0$ at $\kappa_2$. Note that the matrix $G_2$ and $\Phi_{\kappa_2}$ in (48) are indexed by the value of $\kappa_2$. Given (47), We will show that holds the corresponding feasible solutions of (46) and (48) at $\kappa_2 + 1$ exist if feasible solutions of (46) and (48) at $\kappa_2$ exist, which proves (79).

The conditions in (78) indicate that $\grave{\phi}_i(\cdot)$ are orthogonal functions with respect to the weight function $\varpi(\tau) = 1$ over $[-r_2, -r_1]$, Assume $\kappa_2 + 1 \leq \varkappa$ and by the structure of $G_2$ in (51) with (77) and (78), we have

$$G_{2,\kappa_2+1}^\top \grave{\Phi}_{\kappa_2+1} G_{2,\kappa_2+1} = [*]\begin{bmatrix} \grave{\Phi}_{\kappa_2} & \mathbf{0}_{\kappa_2+1} \\ * & \grave{\varphi}_{\kappa_2+1} \end{bmatrix} \begin{bmatrix} G_{2,\kappa_2} \\ \mathbf{g}_{\kappa_2+1}^\top \end{bmatrix} = G_{2,\kappa_2}^\top \grave{\Phi}_{\kappa_2} G_{2,\kappa_2} + \grave{\varphi}_{\kappa_2+1} \mathbf{g}_{\kappa_2+1} \mathbf{g}_{\kappa_2+1}^\top, \tag{80}$$

where $\mathbf{g}_{\kappa_2+1} \in \mathbb{R}^{3+d+\delta}$ can be easily determined by the structure of $G_2$ with (64) and (77), and $G_{2,\kappa_2+1}$ denotes the corresponding $G_2$ at $\kappa_2 + 1$. Note that here that no increase of the dimension indexes $d$, $\delta$, $p_1$ and $p_2$ occurs. By (80) and considering the structure of the matrix inequalities in (46) and (48), we have

$$\begin{aligned} \mathbf{P}_{\kappa_2+1} &= \mathbf{P}_{\kappa_2} + \Pi^\top \left(\grave{\varphi}_{\kappa_2+1} \mathbf{g}_{\kappa_2+1} \mathbf{g}_{\kappa_2+1}^\top \otimes S_2\right) \Pi \\ \widetilde{\boldsymbol{\Omega}}_{\kappa_2+1} &= \widetilde{\boldsymbol{\Omega}}_{\kappa_2} + \left(\mathsf{O}_{q+2\nu} \oplus \Pi^\top \left(\grave{\varphi}_{\kappa_2+1} \mathbf{g}_{\kappa_2+1} \mathbf{g}_{\kappa_2+1}^\top \otimes U_2\right) \Pi \oplus \mathsf{O}_{\mu\nu}\right). \end{aligned} \tag{81}$$

Since $\grave{\varphi}_{\kappa_2+1} > 0$, $\mathbf{g}_{\kappa_2+1}\mathbf{g}_{\kappa_2+1}^\top \succeq 0$ and $S_2 \succeq 0$, $U_2 \succeq 0$ in (47), it is clearly to see that the feasible solutions of $\mathbf{P}_{\kappa_2} \succ 0$, $\widetilde{\mathbf{\Omega}}_{\kappa_2} \succ 0$ infer the existence of a feasible solution of $\mathbf{P}_{\kappa_2+1} \succ 0$, $\widetilde{\mathbf{\Omega}}_{\kappa_2+1} \succ 0$ given the prerequisites of Corollary 3. This finishes the proof. □

**Remark 15.** A hierarchical pattern of the LMIs in Theorem 1 can be also established for the situation when $\acute{\boldsymbol{\phi}}(\cdot)$ contains orthogonal functions which satisfies appropriate constraints resembling to (77) and (78). Note that the corresponding hierarchy result can be derived without using congruence transformations, since the dimensions of $\mathbf{P}$ in (46) and $\widetilde{\mathbf{\Omega}}$ in (48) are not related to the dimensions of $\acute{\boldsymbol{\phi}}(\tau) \in \mathbb{R}^{\kappa_1}$.

On the other hand, a hierarchy of the stability condition in Theorem 1 can be also established with respect to $\grave{\boldsymbol{g}}(\cdot)$ and its dimensions.

**Corollary 4.** *Given the functions with the parameters in* (3)–(12)*, let* $\acute{\boldsymbol{g}}(\cdot)$*,* $\grave{\boldsymbol{g}}(\cdot)$ *and* $N_1$*,* $N_2$ *in Theorem 1 be given where* $\grave{\boldsymbol{g}}(\tau) = \mathsf{Col}_{i=1}^{p_2}\,\grave{g}_i(\tau)$ *with* $\{\grave{g}_i(\cdot)\}_{i=1}^{p_2} \subset \{\grave{g}_i(\cdot)\}_{i=1}^{\infty} \subset \mathbb{C}^1([-r_2,-r_1] \,⨾\, \mathbb{R})$ *satisfying*

$$\exists \alpha \in \mathbb{N},\ \ \forall p_2 \in \{j \in \mathbb{N} : j \leq \alpha\},\ \exists! N_2 \in \mathbb{R}^{p_2 \times \delta},\ \ (r_2+\tau)\frac{\mathsf{d}}{\mathsf{d}\tau}\mathop{\mathsf{Col}}_{i=1}^{p_2}\grave{g}_i(\tau) = N_2\grave{\boldsymbol{g}}(\tau) \tag{82}$$

$$\forall p_2 \in \mathbb{N},\ \ \grave{\mathsf{G}}_{p_2} = \int_{-r_2}^{-r_1}\mathop{\mathsf{Col}}_{i=1}^{p_2}\grave{g}_i(\tau)\mathop{\mathsf{Row}}_{i=1}^{p_2}\grave{g}_i(\tau)\mathsf{d}\tau = \bigoplus_{j=1}^{p_2}\grave{\mathsf{g}}_j,\quad \grave{\mathsf{g}}_j^{-1} = \int_{-r_2}^{-r_1}(\tau+r_2)\grave{g}_j^2(\tau)\mathsf{d}\tau. \tag{83}$$

*Then we have*

$$\forall p_2 \in \{j \in \mathbb{N} : j \leq \alpha\},\ \ \mathcal{H}_{p_2} \subseteq \mathcal{H}_{p_2+1} \tag{84}$$

*where* $\alpha \in \mathbb{N}$ *is given and*

$$\mathcal{H}_{p_2} := \left\{(r_1,r_2)\ \middle|\ r_1 > 0, r_2 > r_1\ \ \&\ \ (46)\text{–}(48)\ \mathit{hold}\ \ \&\ \ P \in \mathbb{S}^l, Q_1;Q_2;R_1;R_2;S_1;S_2;U_1;U_2 \in \mathbb{S}^\nu\right\}$$

*with* $l := n + 2\nu + (d+\delta)\nu$*.*

*Proof.* The proof is similar to the proof of Corollary 3 apart from the fact that for Corollary 4 one only need to consider the increase of the value of $p_2$ instead of $\kappa_2$ in Corollary 3. Given $r_2 > r_1 > 0$ with all the parameters in (3)–(7) and (11) and (12), let $\mathsf{Col}(r_1,r_2) \in \mathcal{H}_{p_2}$ with $\mathcal{H}_{p_2} \neq \varnothing$ which infers that there exist feasible solutions for (46)–(48). Let the dimensions and elements of $\acute{\boldsymbol{f}}(\tau)$, $\grave{\boldsymbol{f}}(\tau)$, $\acute{\boldsymbol{\phi}}(\tau)$, $\grave{\boldsymbol{\phi}}(\tau)$ and $\acute{\boldsymbol{g}}(\tau)$ to be all fixed, and let $P \in \mathbb{S}^l$ and $Q_1;Q_2;R_1;R_2;S_1;S_2;U_1;U_2$ to be a given feasible solution for $\mathbf{P}_{p_2} \succ 0$, (47) and $\widetilde{\mathbf{\Omega}} \prec 0$ at $p_2$. Note that the matrix $H_2$ and $\mathsf{G}_{p_2}$ in (48) are indexed by the value of $\kappa_2$ whereas $\widetilde{\mathbf{\Omega}} \prec 0$ is not related to $\acute{\boldsymbol{g}}(\tau)$ and $\grave{\boldsymbol{g}}(\tau)$ or their dimensions $p_1$, $p_2$. Given (47), we will show that the corresponding feasible solutions of (46) and (48) at $p_2+1$ exist if feasible solutions of (46) and (48) at $p_2$ exist, which leads to (79).

The constraints in (83) show that $\grave{g}_i(\cdot)$ contains functions which are orthogonal with respect to the weight function $\varpi(\tau) = (\tau+r_2)$ over $[-r_2,-r_1]$. Suppose $p_2+1 \leq \alpha$. Now by the structure of $H_2$ in (53) and (82) and (83), we have

$$H_{2,p_2+1}^\top\grave{\mathsf{G}}_{p_2+1}H_{2,p_2+1} = [*]\begin{bmatrix}\grave{\mathsf{G}}_{p_2} & \mathbf{0}_{p_2}\\ * & \grave{\mathsf{g}}_{p_2+1}\end{bmatrix}\begin{bmatrix}H_{2,p_2}\\ \mathbf{h}_{p_2+1}^\top\end{bmatrix} = H_{2,p_2}^\top\grave{\mathsf{G}}_{p_2}H_{2,p_2} + \grave{\mathsf{g}}_{p_2+1}\mathbf{h}_{p_2+1}\mathbf{h}_{p_2+1}^\top \tag{85}$$

where $\mathbf{h}_{p_2+1} \in \mathbb{R}^{3+d+\delta}$ can be easily determined by the structure of $H_2$ with (74) and (82), and $H_{2,p_2+1}$ denotes the corresponding $H_2$ at $p_2+1$. Note that here the values of the dimension indexes $d$, $\delta$, $\kappa_1$, $\kappa_2$ and $p_1$ remain unchanged.

By (85) and considering the structure of $\mathbf{P} \succ 0$ in (46), it yields

$$\mathbf{P}_{p_2+1} = \mathbf{P}_{p_2} + \Pi^\top\left(\grave{\mathsf{g}}_{p_2+1}\mathbf{h}_{p_2+1}\mathbf{h}_{p_2+1}^\top \otimes U_2\right)\Pi. \tag{86}$$

Since $\grave{\mathsf{g}}_{p_2+1} > 0$, $\mathbf{h}_{p_2+1}\mathbf{h}_{p_2+1}^\top \succeq 0$ with $U_2 \succeq 0$ in (47), one can conclude that the feasible solutions of $\mathbf{P}_{p_2} \succ 0$ infer the existence of the feasible solution of $\mathbf{P}_{p_2+1} \succ 0$ given the prerequisites in Corollary 4. On the other hand, since the inequality in (48) is not related to $\grave{\boldsymbol{g}}(\tau)$, thus $\widetilde{\mathbf{\Omega}} \prec 0$ remains unchanged at $p_2+1$. This finishes the proof. □

**Remark 16.** Following the strategy in proving Corollary 4, a hierarchy of conditions in Theorem 1 can be also established when $\acute{\boldsymbol{g}}(\cdot)$ contains orthogonal functions satisfying appropriate constraints resembling (77) and (78). Note that the dimensions of $\mathbf{P}$ in (46) and $\widetilde{\boldsymbol{\Omega}}$ in (48) are not related to the dimensions of $\acute{\boldsymbol{g}}(\tau) \in \mathbb{R}^{p_1}$.

## 5. Numerical examples

In this section, numerical examples are presented to demonstrate the effectiveness of our proposed methods. All examples were tested in Matlab environment using Yalmip Löfberg (2004) with SDPT3 Toh *et al.* (2012) as the numerical solver.

### 5.1. Stability analysis of a distributed delay system

Consider the following distributed delay system

$$\dot{x}(t) = 0.33x(t) - 5\int_{-r}^{0} \sin(\cos(12\tau))x(t+\tau)\mathsf{d}\tau = 0.33x(t) - \begin{bmatrix}5 & \mathbf{0}^\top\end{bmatrix}\int_{-r}^{0}\begin{bmatrix}\varphi_1(\tau)\\ \boldsymbol{f}(\tau)\end{bmatrix}x(t+\tau)\mathsf{d}\tau, \quad t \geq t_0 \tag{87}$$

with any $t_0 \in \mathbb{R}$, where $\varphi_1(\tau) = \sin(\cos(12\tau))$. The corresponding state space matrices of (1) for (87) and (7) are $A_1 = 0.33$ and $A_3 = -\begin{bmatrix}5 & \mathbf{0}^\top\end{bmatrix}$ and the rest of the state space matrices in (1) is zero with $m = q = 0$.

Here we consider two cases for $\boldsymbol{f}(\cdot)$. The first one is $\boldsymbol{f}(\tau) = \boldsymbol{\ell}_d\left(\frac{\tau}{r}\right) = \mathsf{Col}_{i=0}^{d}\, \ell_i\left(\frac{\tau}{r}\right)$ with

$$\ell_d(\tau) := \sum_{k=0}^{d}\binom{d}{k}\binom{d+k}{k}\tau^k \tag{88}$$

containing Legendre polynomials with $\mathsf{F}_1^{-1} = \int_{-r}^{0}\boldsymbol{\ell}_d(\frac{\tau}{r})\boldsymbol{\ell}_d^\top(\frac{\tau}{r})\mathsf{d}\tau = r^{-1}\bigoplus_{i=0}^{d} 2i+1$ and the corresponding $M_1$ in (3) can be easily determined. The second one $\boldsymbol{f}(\tau) = \boldsymbol{h}_d(\tau) = \mathsf{Col}\left[1,\ \mathsf{Col}_{i=1}^{d/2}\sin 12i\tau,\ \mathsf{Col}_{i=1}^{d/2}\cos 12i\tau\right]$ contains trigonometric functions which corresponds to $M_1 = 0 \oplus \begin{bmatrix} \mathsf{O}_{d/2} & \bigoplus_{i=1}^{d/2} 12i \\ -\bigoplus_{i=1}^{d/2} 12i & \mathsf{O}_{d/2}\end{bmatrix}$ satisfying the first relation in (3). Note that $d$ in $\boldsymbol{h}_d(\tau)$ must be positive even numbers and the functions in $\boldsymbol{h}_d(\tau)$ are not orthogonal over $[-r, 0]$ thus the associated $\mathsf{F}$ for $\boldsymbol{h}_d(\tau)$ is not a diagonal matrix. Since $0.33 > 0$, thus the method in Münz *et al.* (2008) cannot be applied. Furthermore, since $\varphi_1(\tau) = \sin(\cos(12\tau))$ does not satisfy the "differentiation closure" property as in (3), the method in Feng & Nguang (2016b) cannot handle (87).

Now apply the spectrum methods in Breda *et al.* (2005) to (87) with $M = 200$. The resulting information of the spectrum of (87) shows that the system is stable in the following intervals: $[0.093, 0.169]$, $[0.617, 0.692]$, $[1.14, 1.216]$, $[1.664, 1.739]$, $[2.188, 2.263]$ and $[2.711, 2.787]$.

In this section we apply a single delay version of Theorem 1 to (87), which is derived via the Krasovskii functional

$$v(\boldsymbol{x}(t), \boldsymbol{y}(t+\cdot)) = \boldsymbol{\eta}^\top(t)P\boldsymbol{\eta}(t) + \int_{-r}^{0}\boldsymbol{y}^\top(t+\tau)\left[Q + (\tau+r)R\right]\boldsymbol{y}(t+\tau)\mathsf{d}\tau \tag{89}$$

as a simplified version of (55), where $P \in \mathbb{S}^{n+(d+1)\nu}$, $Q; R \in \mathbb{S}^{\nu}$ and $\boldsymbol{\eta}(t) := \mathsf{Col}\left[\boldsymbol{x}(t),\ \int_{-r}^{0}F_d(\tau)\boldsymbol{y}(t+\tau)\mathsf{d}\tau\right]$ with $F_d(\tau) = \boldsymbol{f}(\tau) \otimes I_\nu$. Furthermore, the corresponding $\boldsymbol{\vartheta}(t)$ in (20) and (65) is defined as $\boldsymbol{\vartheta}(t) := \mathsf{Col}\left[\boldsymbol{x}(t),\ \boldsymbol{y}(t-r),\ \int_{-r}^{0}F_d(\tau)\boldsymbol{y}(t+\tau)\mathsf{d}\tau\right]$. Now apply the corresponding stability condition derived by (89) with an one delay version congruence transformation (76) with $\eta_1 = 1$ to (87) with $\boldsymbol{f}(\tau) = \boldsymbol{\ell}_d(\frac{\tau}{r})$ and $\boldsymbol{f}(\tau) = \boldsymbol{h}_d(\tau)$, respectively. The results concerning detectable delay margins are summarized in Table 1 and 2. Note that the values of $N$ and $d$ in these tables are presented when the margins of the stable delay intervals can be determined by the numerical results produced by Theorem 1 or the method in Seuret *et al.* (2015). Note that also the results in Table 1 and 2 associated with Breda *et al.* (2005, 2015) are calculated with $M = 200$. Finally, NoDVs in Table 1 and 2 stands for the number of decision variables.

| Breda *et al.* (2005, 2015) | $[0.093, 0.169]$ | $[0.617, 0.692]$ | $[1.14, 1.216]$ |
|---|---|---|---|
| Theorem 1<br>$\boldsymbol{f}(\tau)$ | $d = 3$ (NoDVs: 17)<br>$\boldsymbol{\ell}_d(\frac{\tau}{r})$ | $d = 6$ (NoDVs: 38)<br>$\boldsymbol{h}_d(\tau)$ | $d = 10$ (NoDVs: 80)<br>$\boldsymbol{h}_d(\tau)$ |
| Seuret *et al.* (2015) | $N = 3$ (NoDVs: 17) | $N = 11$ (NoDVs: 93) | $N = 23$ (NoDVs: 327) |

Table 1: Testing of stable delay margins

| Breda *et al.* (2005) | $[1.664, 1.739]$ | $[2.188, 2.263]$ | $[2.711, 2.787]$ |
|---|---|---|---|
| Theorem 1<br>$\boldsymbol{f}(\tau)$ | $d = 10$ (NoDVs: 80)<br>$\boldsymbol{h}_d(\tau)$ | $d = 10$ (NoDVs: 80)<br>$\boldsymbol{h}_d(\tau)$ | $d = 10$ (NoDVs: 80)<br>$\boldsymbol{h}_d(\tau)$ |
| Seuret *et al.* (2015) | − | − | − |

Table 2: Testing of stable delay margins

Note that in Table 1 and 2 the results correspond to Seuret *et al.* (2015) are produced by our Theorem 1 via (89) with $\boldsymbol{f}(\tau) = \boldsymbol{\ell}_d(\frac{\tau}{r})$ and $d = N$ which is essentially equivalent to the method in Seuret *et al.* (2015). With $N = 25$, the margins of the stable delay intervals $[1.664, 1.739]$, $[2.188, 2.263]$ and $[2.711, 2.787]$ still cannot be detected by polynomials approximation approach proposed in Seuret *et al.* (2015). For $N > 25$, our experiments show that the computational time becomes too long to accurately obtain the values of the approximation coefficient and error term via the function `vpaintegral` in Matlab. On the other hand, the function `integral` in Matlab is not an alternative option to calculate the approximation coefficient and error term in this case due to its limited capacity of numerical accuracy. The results in Tables 1 and 2 can be explained by the fact that $\varphi_1(\tau) = \sin(\cos(12\tau))$, $\tau \in [a, b]$ is not "easy" to be approximated by polynomials when the length of $[a, b]$ become relatively large. Consequently, we have shown the the advantage of our method over the one in Seuret *et al.* (2015) when it comes to the stability analysis of (87).

### *5.2. Stability and dissipativity analysis with distributed delays*

Consider a system of the form (1) with $r_1 = 2$, $r_2 = 4.05$ and the state space matrices

$$\begin{aligned}
&A_1 = \begin{bmatrix} 0.01 & 0 \\ 0 & -3 \end{bmatrix}, A_2 = \begin{bmatrix} 0 & 0.1 \\ 0.2 & 0 \end{bmatrix}, A_3 = \begin{bmatrix} -0.1 & 0 \\ 0 & -0.2 \end{bmatrix}, A_6 = I_2, A_7 = A_8 = \mathsf{O}_2,\ D_1 = \begin{bmatrix} 0.2 \\ 0.3 \end{bmatrix} \\
&A_4\left(\begin{bmatrix} \boldsymbol{\varphi}_1(\tau) \\ \acute{\boldsymbol{f}}(\tau) \end{bmatrix} \otimes I_2\right) = \begin{bmatrix} 3\sin(18\tau) & -0.3\mathsf{e}^{\cos(18\tau)} \\ 0 & 3\sin(18\tau) \end{bmatrix},\ A_5\left(\begin{bmatrix} \boldsymbol{\varphi}_2(\tau) \\ \grave{\boldsymbol{f}}(\tau) \end{bmatrix} \otimes I_2\right) = \begin{bmatrix} -10\cos(18\tau) & 0 \\ 0.5\mathsf{e}^{\sin(18\tau)} & -10\cos(18\tau) \end{bmatrix} \\
&C_1 = \begin{bmatrix} -0.1 & 0.2 \\ 0 & 0.1 \end{bmatrix},\ C_2 = \begin{bmatrix} -0.1 & 0 \\ 0 & 0.2 \end{bmatrix},\ C_3 = \begin{bmatrix} 0 & 0.1 \\ -0.1 & 0 \end{bmatrix},\ D_2 = \begin{bmatrix} 0.12 \\ 0.1 \end{bmatrix} \\
&C_4\left(\begin{bmatrix} \boldsymbol{\varphi}_1(\tau) \\ \acute{\boldsymbol{f}}(\tau) \end{bmatrix} \otimes I_\nu\right) = 0.1 \oplus 0,\ C_5\left(\begin{bmatrix} \boldsymbol{\varphi}_2(\tau) \\ \grave{\boldsymbol{f}}(\tau) \end{bmatrix} \otimes I_\nu\right) = 0.2 \oplus 0.1,\ C_6 = \begin{bmatrix} 0 & 0 \\ 0 & 0.1 \end{bmatrix}, C_7 = \begin{bmatrix} 0.2 & 0 \\ 0 & 0 \end{bmatrix}
\end{aligned} \tag{90}$$

with $\boldsymbol{\varphi}_1(\tau) = \boldsymbol{\varphi}_2(\tau) = \begin{bmatrix} \mathsf{e}^{\sin(18\tau)} \\ \mathsf{e}^{\cos(18\tau)} \end{bmatrix}$ and $n = m = 2$, $q = 1$. We find out that the system with (90) is stable by applying the Matlab toolbox of the spectral method in Breda *et al.* (2015). Moreover, the minimization of $\mathbb{L}^2$ gain $\gamma$ is applied as the performance criterion for the system, which corresponds to

$$\gamma > 0,\ \ J_1 = -\gamma I_2,\ \ \widetilde{J} = I_2,\ \ J_2 = \mathbf{0}_2,\ \ J_3 = \gamma \tag{91}$$

in (24).

Even one assumes the method in Münz *et al.* (2009) can be extended to handle systems with multiple delay channels, it still cannot be applied here given that $A_1$ is not a Hurwitz matrix. In addition, since $\boldsymbol{\varphi}_1(\tau) = \boldsymbol{\varphi}_2(\tau)$ does not satisfy the "differentiation closure" property in (3), thus the problem of dissipativity and stability analysis may not be solved by a simple extension of the corresponding conditions in Feng & Nguang (2016b) for a linear CDDS, even a multiple distinct delays version of the method in Feng & Nguang (2016b) is derivable.

Let

$$\acute{\boldsymbol{f}}(\tau) = \acute{\boldsymbol{\phi}}(\tau) = \begin{bmatrix} 1 \\ \mathsf{Col}_{i=1}^{d} \sin 18i\tau \\ \mathsf{Col}_{i=1}^{d} \cos 18i\tau \end{bmatrix}, \quad \grave{\boldsymbol{f}}(\tau) = \grave{\boldsymbol{\phi}}(\tau) = \begin{bmatrix} 1 \\ \mathsf{Col}_{i=1}^{\delta} \sin 18i\tau \\ \mathsf{Col}_{i=1}^{\delta} \cos 18i\tau \end{bmatrix} \tag{92}$$

in (90) and (3), which correspond to

$$M_1 = M_3 = 0 \oplus \begin{bmatrix} \mathsf{O}_d & \bigoplus_{i=1}^{d} 18i \\ -\bigoplus_{i=1}^{d} 18i & \mathsf{O}_d \end{bmatrix}, \quad M_2 = M_4 = 0 \oplus \begin{bmatrix} \mathsf{O}_\delta & \bigoplus_{i=1}^{\delta} 18i \\ -\bigoplus_{i=1}^{\delta} 18i & \mathsf{O}_\delta \end{bmatrix} \tag{93}$$

in (3). Considering $\acute{\boldsymbol{f}}(\cdot)$, $\grave{\boldsymbol{f}}(\cdot)$ in (92) and $\boldsymbol{\varphi}_1(\tau) = \boldsymbol{\varphi}_2(\tau) = \begin{bmatrix} \mathsf{e}^{\sin(18\tau)} \\ \mathsf{e}^{\cos(18\tau)} \end{bmatrix}$ with (14) and (11), we obtain

$$\begin{aligned} A_4 &= \begin{bmatrix} \mathsf{O}_2 & \begin{matrix} 0 & -0.3 \\ 0 & 0 \end{matrix} & \mathsf{O}_2 & \begin{matrix} 3 & 0 \\ 0 & 3 \end{matrix} & \mathsf{O}_{2\times(4d-2)} \end{bmatrix} \\ A_5 &= \begin{bmatrix} \begin{matrix} 0 & 0 & 0 & 0 \\ 0.5 & 0 & 0 & 0 \end{matrix} & \mathsf{O}_{2\times 2\delta+2} & \begin{matrix} -10 & 0 \\ 0 & -10 \end{matrix} & \mathsf{O}_{2\times 2\delta-2} \end{bmatrix} \\ C_4 &= \begin{bmatrix} \mathsf{O}_{2\times 4} & 0.1 \oplus 0 & \mathsf{O}_{2\times 4d} \end{bmatrix}, \quad C_5 = \begin{bmatrix} \mathsf{O}_{2\times 4} & 0.2 \oplus 0.1 & \mathsf{O}_{2\times 4\delta} \end{bmatrix} \end{aligned} \tag{94}$$

which corresponds to the distributed delay terms in (90).

Now apply the conditions (47),(48) and (76)[4] with $\eta_1 = \eta_2 = 1$ and the system's parameters in (90)–(94) where $\acute{\Gamma}_d$, $\grave{\Gamma}_\delta$ are in line with the structure in (13) and the matrices $\acute{\Gamma}_d$, $\grave{\Gamma}_\delta$, $\acute{\mathsf{E}}_d$, $\grave{\mathsf{E}}_\delta$ and $\acute{\mathsf{F}}_d$, $\grave{\mathsf{F}}_\delta$ are calculated computationally via the function `vpaintegral` in Matlab which can produce results with high-numerical precisions. With $d = \delta = 1$ a feasible result can be produced with $\min \gamma = 0.64655$ which requires 196 decision variables. With $d = \delta = 2$, we obtain feasible solutions with $\min \gamma = 0.32346$ requiring 376 variables. Finally, with $d = \delta = 10$ our method can produce feasible solutions with $\min \gamma = 0.31265$ with 4120 variables. It is worthy to mention that even with $d = \delta = 10$ which is a relatively large value, the duration of the calculations of $\acute{\Gamma}_d$, $\grave{\Gamma}_\delta$, $\acute{\mathsf{E}}_d$, $\grave{\mathsf{E}}_\delta$ and $\acute{\mathsf{F}}_d$, $\grave{\mathsf{F}}_\delta$ by `vpaintegral` is still acceptable (about a minute).

On the other hand, let $\acute{\boldsymbol{f}}(\tau) = \boldsymbol{\ell}_d(\frac{\tau}{r})$ and $\grave{\boldsymbol{f}}(\tau) = \boldsymbol{\ell}_d(\frac{\tau+r_1}{r_2-r_1})$ which are Legendre polynomials associated with $\acute{\mathsf{F}}_d = r_1^{-1}\mathsf{D}_d$ and $\grave{\mathsf{F}}_\delta = r_3^{-1}\mathsf{D}_\delta$. The characteristics of the functions in $\boldsymbol{\varphi}_1(\tau) = \boldsymbol{\varphi}_2(\tau)$ indicate that they might be very difficult to be approximated by polynomials. Indeed, let $d = \delta = 15$ with the corresponding $A_4$,$A_5$ and $C_4$,$C_5$. In this case, Theorem 1 with (76) yields no feasible solutions.

## 6. Conclusion

In this paper, a new method for the dissipativity and stability analysis of a linear CDDS with distributed delays in state and output equations has been proposed in Theorem 1 in terms of LMIs. The proposed approach can handle distributed delay with $\mathbb{L}^2$ functions kernel and simultaneously includes approximation errors in the resulting conditions (46)–(48) thanks to the novel integral inequality in (27). In comparison to existing approach in Seuret *et al.* (2015) which depends on the application of Legendre polynomials approximations, the proposed method allows one to apply a broader class of elementary functions $\acute{\boldsymbol{f}}(\cdot)$ and $\grave{\boldsymbol{f}}(\cdot)$ to approximate the distributed delay kernels of (1). Because of the fact that the generality of the Krasovskii functional (55) is also related to the structure of $\acute{\boldsymbol{f}}(\cdot)$ and $\grave{\boldsymbol{f}}(\cdot)$, thus our proposed methods derived from constructing (55) can produce less conservative results compared to a functional parameterized by Legendre polynomials such as the one considered in Seuret *et al.* (2015). The results of numerical examples we have tested have clearly demonstrated the advantage of the proposed methodologies over existing approaches. A potential future direction is to investigate if the hierarchy conclusion in this paper can be derived without having an orthogonality constraint.

[4] Note that here we do not apply (46) in Theorem 1, see Remark 12 for further details.

## Acknowledgement

The authors would like to thank Prof. Keqin Gu for his comments on the choice of the function space for the states of CDDS. In addition, we thank Prof. Johan Löfberg and Prof. Dimitri Breda for the remarks concerning their Matlab software packages.